\documentclass[twocolumn,usenatbib]{mn2e}
\usepackage{graphicx}
\usepackage{natbib}
\usepackage{bm}
\usepackage{amsfonts}
\usepackage{latexsym}
\usepackage[latin1]{inputenc}
\usepackage{amsmath}
\usepackage{epsfig}
\usepackage{amsbsy}
\usepackage{mnfixes}

\newcommand{\mtb}[1]{\mathbf{#1}}

\newcommand{\n}{\mathrm n}
\newcommand{\p}{\mathrm p}
\newcommand{\e}{\mathrm e}
\newcommand{\x}{\mathrm x}

\def \nn  {\nonumber}

\def\jnl@style{\rm}
\def\aaref@jnl#1{{\jnl@style#1}}

\def\aaref@jnl#1{{\jnl@style#1}}

\def\aj{\aaref@jnl{AJ}}                   
\def\apj{\aaref@jnl{ApJ}}                 
\def\apjl{\aaref@jnl{ApJ}}                
\def\apjs{\aaref@jnl{ApJS}}               
\def\apss{\aaref@jnl{Ap\&SS}}             
\def\aap{\aaref@jnl{A\&A}}                
\def\aapr{\aaref@jnl{A\&A~Rev.}}          
\def\aaps{\aaref@jnl{A\&AS}}              
\def\mnras{\aaref@jnl{MNRAS}}             
\def\prd{\aaref@jnl{Phys.~Rev.~D}}        
\def\prl{\aaref@jnl{Phys.~Rev.~Lett.}}    
\def\qjras{\aaref@jnl{QJRAS}}             
\def\skytel{\aaref@jnl{S\&T}}             
\def\ssr{\aaref@jnl{Space~Sci.~Rev.}}     
\def\zap{\aaref@jnl{ZAp}}                 
\def\nat{\aaref@jnl{Nature}}              
\def\aplett{\aaref@jnl{Astrophys.~Lett.}} 
\def\apspr{\aaref@jnl{Astrophys.~Space~Phys.~Res.}} 
\def\physrep{\aaref@jnl{Phys.~Rep.}}      
\def\physscr{\aaref@jnl{Phys.~Scr}}       

\title[The $f$-mode instability in neutron stars]{Nonlinear viscous damping and gravitational wave detectability of the $f$-mode instability in neutron stars}

\author[A. Passamonti  $\&$ K. Glampedakis ]
{Andrea Passamonti$^1$ $\&$ Kostas Glampedakis$^{1,2}$ \\ \\
$^1$ Theoretical Astrophysics, University of T\"{u}bingen, Auf der
Morgenstelle 10, T\"{u}bingen D-72076, Germany \\
$^2$ Departamento de F\'{i}sica, Universidad de Murcia, E-30100 Murcia, Spain}

\begin{document}

\date{\today}

\pagerange{\pageref{firstpage}--\pageref{lastpage}} \pubyear{}
\maketitle

\label{firstpage}


\begin{abstract}
We study the damping of the gravitational radiation-driven $f$-mode instability in rotating neutron stars by nonlinear bulk viscosity 
in the so-called supra-thermal regime. In this regime the dissipative action of bulk viscosity is known to be enhanced as a result of 
nonlinear contributions with respect to the oscillation amplitude. Our analysis of the $f$-mode instability is based on
a time-domain code that evolves linear perturbations of rapidly rotating polytropic neutron star models. The extracted 
mode frequency and eigenfunctions are subsequently used in standard energy integrals for the gravitational wave 
growth and viscous damping. We find that nonlinear bulk viscosity has a moderate impact on the size of the $f$-mode instability window, 
becoming an important factor and saturating the mode's growth at a relatively large oscillation amplitude. 
We show similarly that nonlinear bulk viscosity leads to a rather high saturation amplitude even for the $r$-mode instability.
In addition, we show that the action of bulk viscosity can be 
significantly mitigated by the presence of superfluidity in neutron star matter.  
 Apart from revising the $f$-mode's instability window we provide results on the mode's gravitational wave detectability.
Considering an $f$-mode-unstable neutron star located in the Virgo cluster and assuming a mode amplitude at the level
allowed by bulk viscosity, we find that the emitted gravitational wave signal could be detectable by advanced ground-based detectors 
such as Advanced LIGO/Virgo and the Einstein Telescope.

\end{abstract}


\section{Introduction}

Neutron stars are exciting cosmic laboratories of matter at extreme conditions. However, unveiling the physics of these ultra-dense and 
relativistic systems is a real challenge even after more than four decades of astronomical observations. 
The situation is likely to improve significantly with the
advent of gravitational wave astronomy. The forthcoming generation of improved gravitational-wave detectors, such as
the Advanced LIGO and Virgo interferometers and the planned Einstein Telescope (ET),  should be sensitive to various  
 aspects of neutron star dynamics. The observation of neutron star oscillations (normal modes), usually dubbed 
neutron star ``asteroseismology'', could become an excellent tool for probing the interior properties of these
objects~\citep{1998MNRAS.299.1059A}.

Of particular importance for neutron star asteroseismology are the oscillation modes that can become unstable 
(i.e. grow in amplitude) under the emission of gravitational radiation.
This is the well-known CFS instability which develops 
when a retrogade mode (with respect to the stellar rotation) becomes prograde, as a result of rotational dragging,  
and then grows while emitting gravitational radiation~\citep{1970PhRvL..24..611C,1978ApJ...222..281F}. 
Exhaustive work on the subject has shown that among the various stellar modes the ones that could become CFS-unstable in realistic 
neutrons stars, and likely to be strong sources of gravitational waves, are the inertial $r$-mode and the fundamental $f$-mode  
(see~\citet{2003CQGra..20R.105A} for a review).

The study of dissipative mechanisms that could limit or entirely suppress the gravitational wave instability of the $r$- and $f$-modes 
has become a small industry since the early days of this field~\citep{2001IJMPD..10..381A}. Given that astrophysical neutron stars
are viscous systems, a clear understanding of the various dissipative mechanisms is a necessary ingredient of any realistic
theoretical model. In fact, this is another facet of neutron star asteroseismology as the absence of unstable modes can be related 
to the strength of dissipation, thus providing indirect information about the stellar interior. 
For instance, this possibility has been recently illustrated by \citet*{2011PhRvL.107j1101H}, who compared  the latest theoretical 
parameter space of the $r$-mode instability against the observed population of accreting  neutron stars in low mass X-ray binaries.

The present paper makes another contribution to our understanding of the $f$-mode instability and 
its damping. We focus on the action of bulk viscosity, which accounts for the energy drained from the
oscillating fluid as a result of the induced departure from local beta equilibrium~\citep{1989PhRvD..39.3804S, 1983bhwd.book.....S}. 
The novelty of our work is the study of bulk viscosity in the {\em nonlinear} regime (with respect to the oscillation amplitude), 
thus extending previous models based on amplitude-independent viscosity~\citep[e.g.][]{1991ApJ...373..213I,1995ApJ...438..265L, 2011PhRvL.107j1102G}.

The bulk viscosity formalism used in our analysis is based on the work by Alford and 
collaborators~\citep{2010JPhG...37l5202A,2011AIPC.1343..580A,2011arXiv1103.3521A}. 
These authors have recently studied the role of nonlinear bulk viscosity in the context of the $r$-mode instability (we also briefly discuss 
the $r$-mode below in Sec~\ref{sec:Rmode}). A key result in these papers is an analytical expression for the bulk viscosity coefficient
in the nonlinear regime which can be easily incorporated in the standard machinery for calculating viscous mode damping. 
This latter procedure consists of solving the inviscid hydrodynamics equations and determining the mode frequency and eigenfunctions, 
and subsequently using these results in the available expressions for the mode's volume-integrated energy and damping rate 
(see section~\ref{sec:Form}).

Our $f$-mode calculation is numerical and is based on a 2D time-domain code developed by~\citet{2009MNRAS.394..730P}. The code 
evolves the linearised hydrodynamical equations of rapidly rotating stars in Newtonian theory. The frequency and eigenfunctions 
of specific modes can be then extracted from the time-evolved perturbations with a method developed by~\citet{Stergioulas:2003ep}
and~\citet{Dimmelmeier:2005zk}. 
An overview and a sample test of the code is provided in section~\ref{sec:NSmodel}.
In this work we focus on the sextupole ($l=m=4$) $f$-mode, which is known to be the dominant 
unstable multipole both in Newtonian and Relativistic stars~\citep{1983PhRvL..51...11F, 1991ApJ...373..213I,2011PhRvL.107j1102G}.
We calculate the resulting instability ``window'' (that is, the unstable portion of the parameter space described by the stellar rotational 
frequency and temperature) after accounting for the action of shear and bulk viscosity (section~\ref{sec:InsWin}). 
Our calculation provides results for the $f$-mode {\em saturation} amplitude as determined by the balance between 
nonlinear viscous damping and gravitational radiation-driven growth (section~\ref{sec:InsWin}). 
We also calculate the gravitational wave strain associated with an unstable $f$-mode and with its amplitude limited
by nonlinear bulk viscosity (section~\ref{sec:GW}). Our conclusions and a discussion of remaining issues can 
be found in section~\ref{sec:concl}.

\section{Formalism} \label{sec:Form}

The strategy for calculating the viscous damping and the gravitational wave-driven growth of oscillation modes is
a well established two-step procedure~\citep{1991ApJ...373..213I} and therefore it will only be outlined here. 
The first step consists of calculating the mode properties (frequency, eigenfunctions) of the inviscid system. In the second step, these 
results are inserted in the volume integral expressions for the damping/growth rates.  This procedure relies on 
the implicit assumption that the mode evolves quasi-adiabatically and the dissipation and growth rates are much lower than the 
oscillation frequency. 

The relevant linearised hydrodynamics equations for non-superfluid matter are the familiar Euler, mass conservation and Poisson equations. In 
a frame rotating with the stellar angular frequency $\Omega^i = \Omega \hat{z}^i$ these are:
\begin{eqnarray}
&& \partial_t \delta v^i + 2 \epsilon^{ijk} \Omega_j \delta v_k 
+ \nabla^i \left( \delta h + \delta \Phi \right) =  \frac{1}{\rho} D^i\, ,                       
\label{eq:dvdt} \\
\nonumber \\
&& \partial_t \delta \rho + \nabla_j ( \rho \delta v^j ) = 0 \, ,      
\label{eq:drdt} \\
\nonumber \\
&& \nabla^2 \delta \Phi =  4 \pi G \, \delta \rho \, ,  \label{eq:dPhi}  
\end{eqnarray}
where $\delta v^i$, $\delta {\bf \rho}$ and $\delta \Phi$ are, respectively, the perturbed velocity, mass density and 
gravitational potential. We have also introduced the perturbation of the specific enthalpy, $\delta h = \delta P /  \rho$, 
where $\delta P$ is the fluid pressure perturbation. The term $D^i$ in the Euler equation~(\ref{eq:dvdt}) is 
the collective viscous force and will be discussed below.

The system of equations~(\ref{eq:dvdt})-(\ref{eq:dPhi}) is closed once an equation of state (EoS) is 
provided. In this work we consider a polytropic EoS
\begin{equation}
P = K \rho ^{\gamma } \, , \label{eq:polEoS}
\end{equation}
where $K$ is a constant and $\gamma$ is related to the adiabatic index by $N =  1 / \left( \gamma -1 \right)$.

Within the present single-fluid hydrodynamics model the viscosity force consists of shear and bulk viscosity
contributions~\citep{1991ApJ...373..213I},
\begin{equation}
D^i =  \nabla_j ( \nu \delta \sigma^{ij}) + \nabla^i (\zeta \delta \sigma )  \, ,
\label{visc1}
\end{equation}
where $\eta$ and $\zeta$ are, respectively, the shear and bulk viscosity coefficient and
\begin{align}
 \delta \sigma^{ij} & = \frac{1}{2} \left(   \nabla^{i} \delta v^{j} +  \nabla^{j} \delta v^{i}    
- \frac{2}{3} g^{ij} \nabla \delta \sigma  \right) \, , \\ 
\delta \sigma  & =  \nabla_j \delta v^j  \, ,
\label{visc2}
\end{align}
where  $g_{ij} $ is the metric tensor.

The viscosity force leads to the following dissipation rates~\citep{1991ApJ...373..213I},
\begin{equation}
\partial_t E_s =  2 \int dV \eta \, \delta \sigma^{ij} \delta \sigma_{ij}^{*} \, ,  \qquad
\partial_t E_b = \int dV  \zeta \, \delta \sigma  \delta \sigma^{*}  \, , \label{eq:dEbdt}
\end{equation}
where the integrals are taken over the stellar volume and an asterisk denotes a complex conjugate. 
The gravitational radiation power is given by the standard multipole formula~\citep{1980RvMP...52..299T}, 
\begin{equation}
\partial_t E_{\rm gw} =  \omega \sum_{l\ge2} N_{l} \left( \omega - m \Omega \right) ^{2 l+1} \left( \left| \delta D_{lm}  
 \right|^2 + \left| \delta J_{lm}   \right|^2\right) 
\, , \label{eq:dEdtgw}
\end{equation}
where $\omega$ is the mode frequency as measured in the rotating frame. This expression features mass
and current multipoles, denoted as $\delta D_{lm} $ and $\delta J_{lm}$ respectively, and the coupling constant
\begin{equation}
N_l = \frac{ 4\pi G }{c^{2l+1}}  \frac{ \left(l+1\right) \left(l+2\right)  }
{ l \left( l -1\right) \left[ \left( 2 l + 1 \right) !! \right] ^2  } \, .
\end{equation}
The $f$-mode oscillation radiates mainly through the mass multipole moments. These are given by
\begin{equation}
\delta D_{lm} = \int dV r^l \delta \rho \, Y_{lm}^{*} \, ,  \label{eq:Dlm} 
\end{equation}
where $Y_{lm}$ is the standard spherical harmonic function with indices $(l,m)$.

The above damping/growth rates can be readily calculated once the mode solution has been obtained from the inviscid
Euler equation, i.e. equation~(\ref{eq:dvdt}) with $D^i=0$. The inviscid equations also lead to the conserved
mode energy
\begin{equation}
E  = \frac{1}{2} \int dV \left[  \rho \, \delta v^{i} \delta v_{i}^{*} + \frac{1}{2} 
\left(  \delta \rho \, \delta U^{*} + \delta \rho^{*} \delta U   \right)  \right] 
\label{eq:En}  \,  ,
\end{equation}
where  $\delta U =\delta h + \delta \Phi$. 

With viscosity included, the mode acquires a complex-valued frequency. For example, the perturbed velocity has a time profile 
\begin{equation} 
\delta v^i  \sim e^{i \omega t - t/\tau } \, ,  
\label{eq:dQ}
\end{equation}
and similarly for the other parameters. As a result  the mode energy evolves according to  
\begin{equation}
\partial_t E = - \frac{2 E}{ \tau} \, .
\end{equation}
The net combined effect of damping and growth can then be expressed as~\citep{1991ApJ...373..213I}
\begin{equation}
\frac{1}{\tau} = \frac{1}{\tau_{\rm gw}} + \frac{1}{\tau_{\rm s}} + \frac{1}{\tau_{\rm b}}, 
\label{eq:tauT} 
\end{equation}
where
\begin{equation}
\frac{1}{\tau_{\rm gw}} =  \frac{\partial_t E_{\rm gw}}{2E},\qquad
\frac{1}{\tau_{\rm s}}  = \frac{\partial_t E_{\rm s}}{2E}, \qquad 
\frac{1}{\tau_{\rm b}}  = \frac{\partial_t E_{\rm b}}{2E}  \, , 
\label{eq:taudef}
\end{equation}
are the timescales associated with, respectively, gravitational wave growth, shear and bulk viscosity damping. 
A mode goes CFS-unstable when $\tau < 0$, which requires, as a necessary condition, a negative flux $\partial_t E_{\rm gw} < 0$.
This can happen only when $\omega \left( \omega - m \Omega \right) \le 0$, which is the mathematical statement of the 
mode changing from retrograde to prograde.

Up to this point our discussion has been essentially a repetition of well known theory. The following sections, however,
offer a new analysis of the effect of bulk viscosity on the $f$-mode instability. On the other hand there is nothing
new here about shear viscosity, which anyway plays a minor role in our calculation. This is because shear viscosity has
a significant impact on the instability in the low temperature regime ($ T \ll 10^9\,\mbox{K}$) where the instability
is anyway likely to be suppressed by superfluid vortex mutual friction~\citep{1995ApJ...444..804L}. 
Our modelling does not account for superfluidity. 
Without superfluidity the shear viscosity is dominated by neutron collisions 
and the resulting coefficient is 
\citep{1979ApJ...230..847F,1991ApJ...373..213I}
\begin{equation}
\eta = 347  \rho^{9/4}  \, T^{-2} \,  \textrm{g cm}^{-1} \textrm{s}^{-1}  \, .  \label{eq:eta}
\end{equation}
We use this coefficient for the shear viscosity in equations~(\ref{eq:dEbdt}) and~(\ref{eq:taudef}).

\subsection{Nonlinear bulk viscosity}
\label{sec:bulk}

Considering a neutron star core composed of $npe$ matter, the beta equilibrium is established through the modified Urca 
(mUrca) reactions
\begin{equation}
\tilde N + n   \to  \tilde N + p + e^{-} + \bar\nu_{e}  \, , \qquad \tilde N + p + e^{-}   \to \tilde N + n + \nu_{e}  \, ,  
\label{eq:mUrca}
\end{equation}
where $\tilde N$ is a spectator nucleon that guarantees overall energy-momentum conservation. 
In more massive neutron stars where the proton fraction is sufficiently large, $\x_{\p} >  1/9 $, the more efficient 
direct Urca (dUrca) process (i.e. the reaction~(\ref{eq:mUrca})  without $\tilde N$) becomes energetically favourable~\citep{1991PhRvL..66.2701L}. Although both processes 
are discussed in the following, most attention is given to the mUrca case which is the relevant one in a `canonical'
neutron star.

The coefficient of the bulk viscosity associated with beta reactions takes the form~\citep{2001A&A...372..130H}
\begin{equation}
\zeta_{ 0} = \frac{C^2 \Gamma T^{2q} }{\omega^2} \, ,
\label{zeta_lin}
\end{equation}
where $\omega$ is the oscillation frequency in the rotating frame, and $\Gamma$ represents the particle reaction rate 
(see Table~\ref{tab:1}). The temperature power-law is reaction-dependent and takes the value $q=3$ ($q=2$) for mUrca (dUrca). 
The quantity $C$ is the so-called strong interaction susceptibility 
\begin{equation}
C \equiv \rho \left. \frac{\partial \mu_{\rm \Delta} }{\partial \rho} \right| _{x_\p} \, ,
\end{equation}
where $\mu_{\rm \Delta} = \mu_{\n} - \mu_{\p} - \mu_{\e}$ is the difference between the particle chemical potentials on the two sides of (\ref{eq:mUrca}). For the simple case of neutron matter modelled as a gas of free (non-interacting) hadrons
we have in natural units~\citep{2010JPhG...37l5202A}  
\begin{equation}
C = \frac{ \left(  3 \pi^2 \rho \right)^{2/3}}  {6 \, m_{\n}^{5/3} } \, ,
\qquad 
\label{eq:C}
\end{equation}
where $m_\n$ is the neutron mass. The resulting viscosity coefficient for this type of matter has been calculated
by~\citet{1989PhRvD..39.3804S} and is the most widely used coefficient in the literature:
\begin{equation}
\zeta_{0} = 6 \times 10^{-59}   \rho^2 \,  \omega^{-2} \, T^6 \,  \textrm{g cm}^{-1} \textrm{s}^{-1} \,  .   
\label{eq:saw} 
\end{equation}
Strictly speaking, the bulk viscosity formulae (\ref{zeta_lin}) and (\ref{eq:saw}) are valid for a `small' oscillation
amplitude  in  the so-called sub-thermal regime, where $\mu _{\Delta} \ll k_{B} T$. However, if the oscillation amplitude is 
sufficiently large the system may enter the supra-thermal regime  $\mu _{\Delta} \gtrsim k_{B} T$. 
As discussed in \citet{2010JPhG...37l5202A, 2011AIPC.1343..580A}, this condition may be actually reached at relatively small 
amplitudes $\Delta \rho / \rho \gtrsim 0.01$. The Lagrangian density perturbation $\Delta \rho$ is defined 
in terms of the displacement vector $\xi^{i}$ as $\Delta \rho = \delta \rho + \xi^{i} \nabla_{i} \rho$.

In this {\em nonlinear} bulk viscosity regime, the formula (\ref{zeta_lin}) is replaced by the amplitude-dependent expression 
\citep{2010JPhG...37l5202A, 2011AIPC.1343..580A},
\begin{equation}
\zeta  =  \zeta_{0} \left[  1 + \sum_{j=1}^{q} \frac{ \left(2j+1 \right)!! \chi_{j} }{  2^j \left( j + 1 \right) ! }  
\left(  \frac{C}{T} \frac{\Delta \rho}{\rho} \right)^{2j} \right] \, ,  
\label{eq:zetaAl}
\end{equation}
where $\chi_j$ are matter/reaction-dependent parameters given in Table~\ref{tab:1}. This expression is approximate, but remains 
accurate as long as the amplitude is not too large. 
For mUrca (dUrca) bulk viscosity the relevant amplitude threshold is $\Delta \rho / \rho \lesssim 1$ 
($\Delta \rho / \rho \lesssim 0.1$)~\citep{2010JPhG...37l5202A, 2011AIPC.1343..580A}.  
For an oscillation amplitude above these thresholds equation~(\ref{eq:zetaAl}) becomes inaccurate, failing to
capture the nonlinear saturation of the supra-thermal bulk viscosity and the subsequent decreasing behaviour 
\citep[see figure 1 in][]{2011AIPC.1343..580A}. For the purpose of this paper, where the fluid perturbation remains in the linear regime 
(i.e. $\Delta \rho / \rho < 1$), the approximation (\ref{eq:zetaAl}) remains accurate and, in fact, includes the 
maximum viscosity attained in the supra-thermal regime.

The nonlinear coefficient (\ref{eq:zetaAl}) provides the main microphysical input for our $f$-mode instability calculation. 
A large amplitude $f$-mode is likely to `activate' the nonlinear terms in (\ref{eq:zetaAl}) and suffer an enhanced viscous 
dissipation. If the resulting damping rate is sufficiently high then the mode may actually saturate and stop growing at some 
maximum amplitude. In the following sections we explore to what extent this scenario could be relevant for unstable $f$-modes of 
rapidly rotating neutron stars. 

In the following, $\zeta$ is calculated assuming the simple case of a free hadron gas, see Table~\ref{tab:1} for the
relevant parameters. For the proton fraction (which is assumed uniform throughout the core)  we take a typical value $x_\p = 0.05$ 
when mUrca is active and a higher value $x_{\p} = 1/9$ in the case of dUrca viscosity.

Given the functional form of $\zeta$ it is also necessary to define an appropriate 
mode-amplitude.  We  introduce the amplitude parameter $\alpha$ defined by the following expression:
\begin{equation}
\delta v^{r} _{eq} = \alpha \, \omega R_{eq}  Y_{lm} \left(\pi/2 , 0 \right)  \, , \label{eq:norm}
\end{equation}
where $R_{eq}$ is the equatorial radius of the star, and $\delta v^{r} _{eq} $ is the radial component of the 
velocity perturbation determined at $r=R_{eq}$ and $\left( \theta, \phi \right) = \left( \pi/2 , 0\right)$. 
Note that our definition of the mode amplitude is similar to the $r$-mode definition used by~\citet{1998PhRvD..58h4020O}, 
provided that $\omega$ is replaced by $\Omega$.

\begin{table}
\begin{center}
  \caption{This table provides the parameters appearing in the expression~(\ref{eq:zetaAl}) 
  for the the bulk viscosity coefficient $\zeta$. Note that these quantities are given in natural units~\citep[see][]{2010JPhG...37l5202A}.
  \label{tab:1}   }
\begin{tabular}{ c c c c c c }
  \hline
 Process & $q$ &  $ \tilde \Gamma  \,  [\textrm{MeV}^{3-2q}] $  & \hspace{-0.35cm} $\chi_{1}^{-1}$  & \hspace{-0.35cm}$\chi_{2}^{-1}$ &  \hspace{-0.35cm}$\chi_{3}^{-1}$ \\ 
  \hline
  mUrca    &  3  &  $4.68\times 10^{-19} \left( \frac{x_{\p} n}{n_0} \right)^{1/3} $ 
  & \hspace{-0.35cm} $ \frac{367}{189}\pi^2 $   & \hspace{-0.35cm} $\frac{367}{21}\pi^4$  & \hspace{-0.35cm}$\frac{1835}{3} \pi^6 $\\ \\
  dUrca    &  2  &  $5.24\times 10^{-15} \left( \frac{x_{\p} n}{n_0} \right)^{1/3} $ 
  & \hspace{-0.35cm} $ \frac{17}{10}\pi^2$    \hspace{-0.35cm}& $17\pi^4$  &  \hspace{-0.35cm}0 \\
  \hline
\end{tabular}
\end{center}
\end{table}

\section{Neutron star model} \label{sec:NSmodel}

Our analysis is based on a Newtonian polytropic stellar model with uniform rotation. More specifically, we study two polytropes 
with indices $N=1$ and $N=3/4$, which provide a reasonable approximation to realistic equations of state. 
The axisymmetric equilibrium configurations are determined by solving the stationary equations with the self-consistent method 
developed by~\cite{1986ApJS...61..479H}. A detailed discussion of the numerical techniques can be found 
in~\cite{2009MNRAS.394..730P, 2009MNRAS.396..951P}. In the following calculations we fix the stellar mass at $M=1.5~M_{\odot}$
and the constant $K$ (in cgs units) at $K=6.637\times10^4$ ($K=1.3346$) for the $N=1$ ($N=3/4$) model. 
The resulting stellar radius of the nonrotating model is then $R = 12.533$~km ($R = 14.245$~km) for $N=1$ ($N=3/4$). 
As discussed by~\citet{1991ApJ...373..213I} and~\citet{1995ApJ...438..265L}, the $f$-mode damping/growth timescale in neutron stars 
with different mass and radius can be obtained by a simple rescaling.

Although our $f$-mode calculation assumes a fluid star we implicitly account for the presence of a solid crust 
by limiting the action of bulk viscosity in the liquid core, where the reactions~(\ref{eq:mUrca}) take place. 
This means that the volume integral in equation~(\ref{eq:dEbdt}) is taken only over
the core. This truncation should be accurate below the crust melting temperature
$T_{\rm melt} \sim (5-7) \times 10^9\,\mbox{K}$~\citep{1983bhwd.book.....S}, a region which overlaps with the temperature range where the 
$f$-mode instability is likely to operate. In our model the ``crust'' occupies the outer layer of the star with its boundary located at the
transition density $\rho_{cc} = 1.2845\times10^{14}~\textrm{g cm}^{-3}$.

A step of the calculation that deserves some special discussion concerns the constraints imposed on the $f$-mode amplitude
by the chosen stellar models. A first requirement is that the mode velocity in the inertial frame should not exceed the speed of light.  
In the rotation range relevant for the instability of the $l=m=4$ $f$-mode (which is the most unstable multipole), 
we find that this condition is satisfied when $\alpha < 1.2 $  ($\alpha< 1.4$) for the $N=1$ ($N=3/4$) model.

Another physically motivated (albeit less rigorous) condition is imposed by the linear perturbation formalism used here. 
We can quantify this requirement in terms of the Lagrangian density perturbation $\Delta \rho$, as  our perturbation formalism 
(section~\ref{sec:Form}) should break down when $\Delta \rho /\rho \gtrsim 1$. This condition also limits  the use of the 
approximate bulk viscosity coefficient given in equation~(\ref{eq:zetaAl}). Indeed, as detailed in~\citet{2011AIPC.1343..580A}, 
this approximation  for the mUrca (dUrca) bulk viscosity is accurate only when $\Delta \rho / \rho \lesssim 1$ ($\Delta \rho / \rho \lesssim 0.1$).  
In other words, we need to make sure that $\Delta \rho/\rho$ remains within the limits of linearised theory.

\begin{figure*}
\begin{center}
\includegraphics[height=74mm]{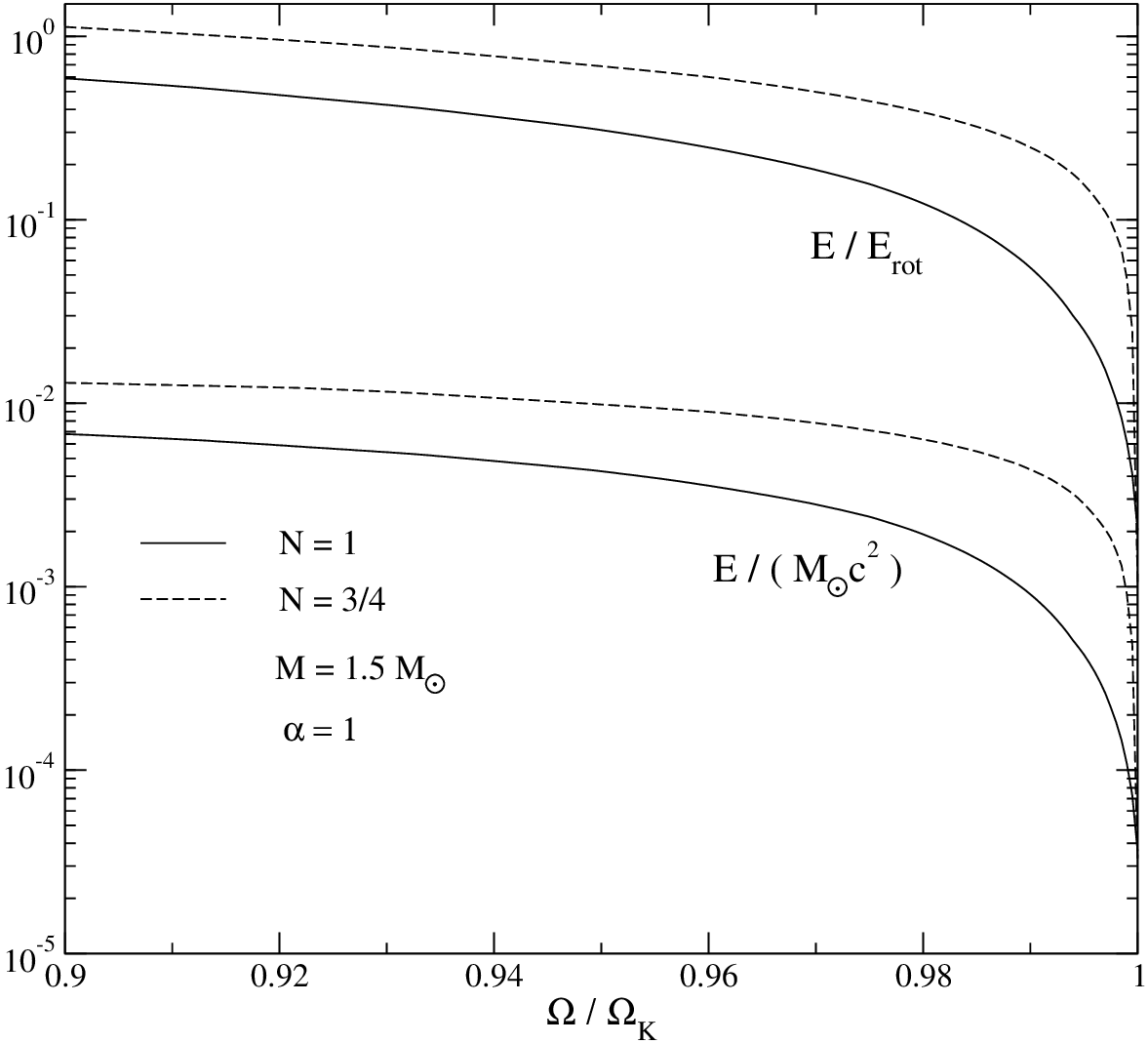}
\hspace{0.2cm}
\includegraphics[height=75mm]{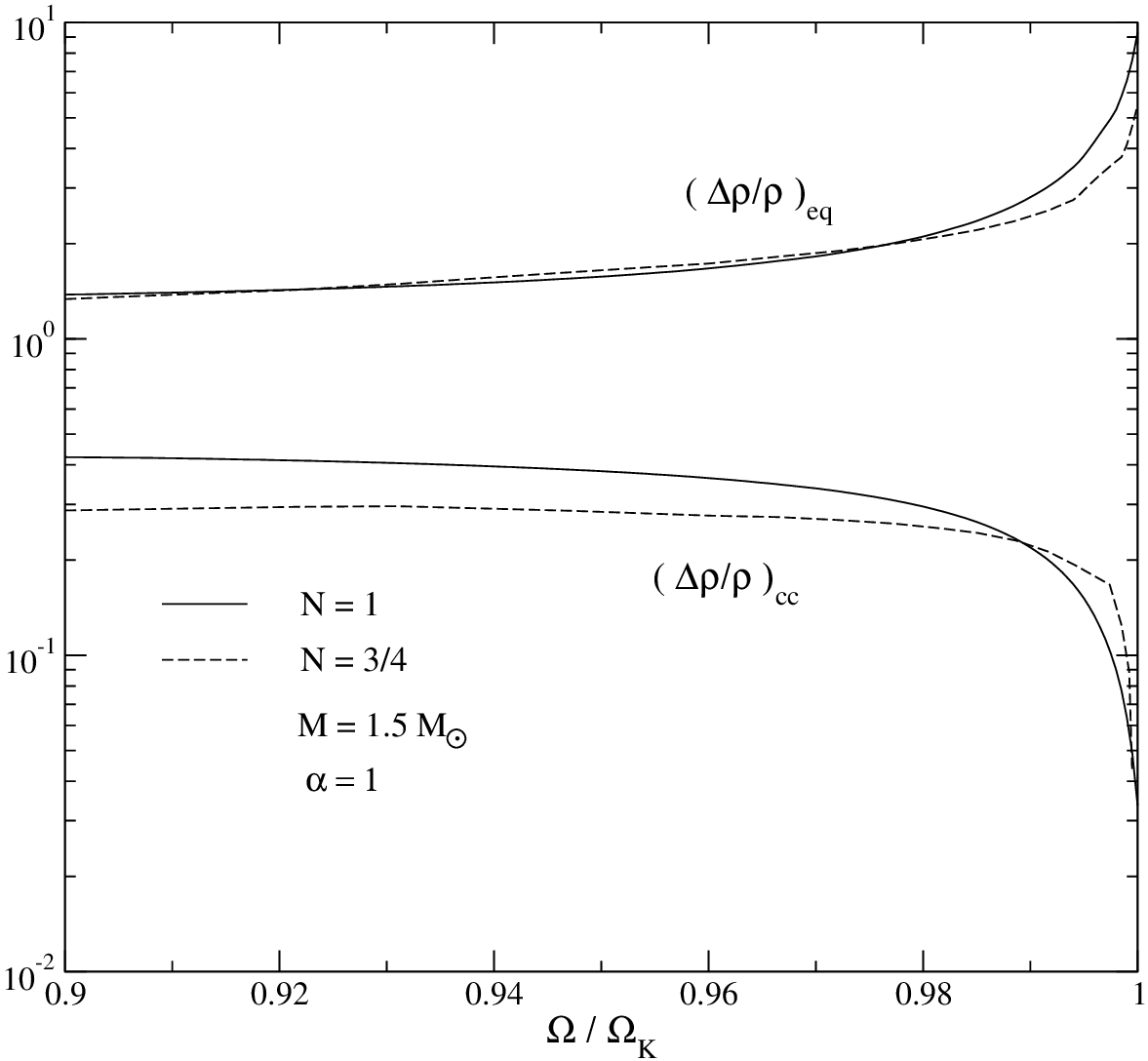}
\caption{The energy  (left panel) and the 
amplitude of $\Delta \rho / \rho $ (right panel) of the $l=m=4$ $f$-mode
in  rapidly rotating models with $\Omega \ge 0.9~\Omega_{\rm K}$.
The stars have a constant mass $M=1.5~M_{\odot}$ and polytropic index $N=1$ and $N=3/4$. In all cases, the mode  amplitude  is $\alpha=1$. 
The density perturbation $\Delta \rho / \rho$  is computed on the equatorial plane ($\theta = \pi / 2$) at two
different radii,  at the bottom of the crust ($r=R_{cc}$) and at the star's surface ($r=R_{eq}$). 
 \label{fig1}}
\end{center}
\end{figure*}

The quantity $\Delta \rho / \rho$ is shown in figure~\ref{fig1} for the  $l=m=4$ $f$-mode with a fiducial amplitude $\alpha=1$. 
The density eigenfunction is calculated 
at the equatorial plane ($\theta=\pi/2$), and in  two radial locations, namely, the surface  and the crust-core interface. 
All  the rapidly rotating models shown in figure~\ref{fig1} have $\Delta \rho / \rho > 1 $ at the surface. 
 On the other hand,  the perturbation remains `well-behaved' at the crust-core boundary, i.e.
$\Delta \rho / \rho < 1 $. The same is true for every point inside the stellar core. Therefore our analysis can be safely used for
studying the bulk viscosity in the core.

Another way of gauging the size of the mode amplitude is by comparing the mode  energy $E$ defined in equation~(\ref{eq:En}) against the
bulk rotational energy of the star $E_{\rm{rot}}$. It is reasonable to expect that in the instability regime we have $E < E_{\rm rot}$, as 
 the latter energy is the one tapped by the instability. 
In figure~\ref{fig1} we show these two quantities for an $f$-mode with $\alpha=1$. Near the maximum rotation rate, 
the mode energy is much smaller than the rotational energy, $E \simeq (10^{-2}-10^{-3})\,E_{\rm{rot}}$, but becomes a significant fraction
of it at relatively lower rotations. For instance, the condition $E < E_{\rm{rot}}$ is violated in the  $N=3/4$ model
when $\Omega \le 0.91 \Omega_{\rm K}$. However this matters little for our calculation since, as we will see in section~\ref{sec:InsWin}, 
these stars lie outside the instability window of the $f$-mode.
In summary,  our $f$-mode calculations preserve self-consistency with respect to the mode amplitude (\ref{eq:norm}) and the 
approximate supra-thermal bulk viscosity coefficient (\ref{eq:zetaAl}).

\subsection{Numerical Code}

The evolution of the perturbation equations (\ref{eq:dvdt})-(\ref{eq:dPhi}) is a three-dimensional problem in space. In order to have an easier 
numerical implementation, we can exploit the background symmetry and expand the non-axisymmetric perturbations in 
a Fourier series with respect to the azimuthal harmonic index~$m$. As a result,  the problem becomes 
 two-dimensional~\citep{1980MNRAS.190...43P, 2009MNRAS.394..730P, 2009MNRAS.396..951P}.  
For the mass density perturbation the expansion is given by
\begin{equation}
\delta \rho \left( t,r,\theta,\phi \right) = \sum_{m=0}^{m=\infty}
               \left[ \delta \rho_{m}^{+} \left( t,r,\theta\right)
               \cos m \phi + \delta \rho_{m}^{-} \left(
               t,r,\theta\right) \sin m \phi \right] \,  , 
               \label{eq:dPexp}
\end{equation}
and similarly for  the other variables.
For any $m$, we evolve a system of ten partial differential
equations for the ten variables $\left(  \delta  v_{i} ^{\pm}, \delta
\rho^{\pm}, \delta \Phi^{\pm} \right)$,  while the enthalpy 
perturbation, $\delta h^{\pm}$, is determined by an EoS. For a polytropic star  it is given by
\begin{equation}
\delta h = c_s^2 \frac{\delta \rho}{\rho} \, ,
\end{equation}
where $c_s^2$ is the speed of sound,  
\begin{equation}
 c_s^2 = \frac{\partial P }{\partial \rho } = K \gamma \rho ^{\gamma-1}   \, .
\end{equation}

The linearised equations are evolved numerically with a code developed 
by~\citet{2002MNRAS.334..933J} and \cite{2009MNRAS.394..730P, 2009MNRAS.396..951P}, where the reader can find all the details.
From the time evolution of each perturbation variable, we determine the mode frequency with 
a Fast Fourier Transformation (FFT) and then extract the eigenfunctions with the method developed by~\cite{Stergioulas:2003ep}
and~\cite{Dimmelmeier:2005zk}. Both the frequencies and eigenfunctions are used in section~\ref{sec:InsWin} to determine the 
damping time of the bulk and shear viscosity and the growth time of the gravitational-wave-driven instability. 

\begin{figure*}
\begin{center}
\includegraphics[height=75mm]{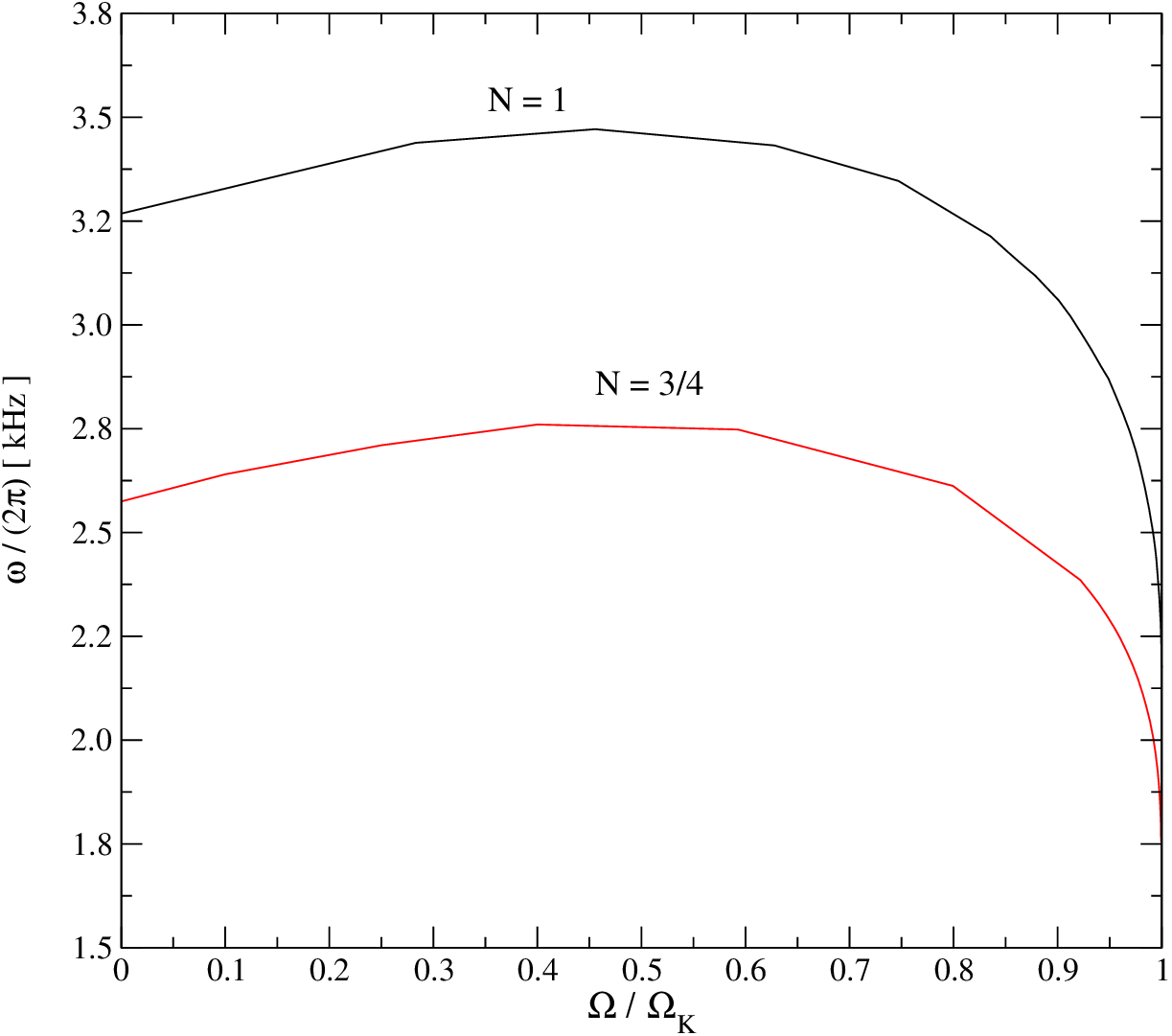}
\hspace{0.2cm}
\includegraphics[height=75mm]{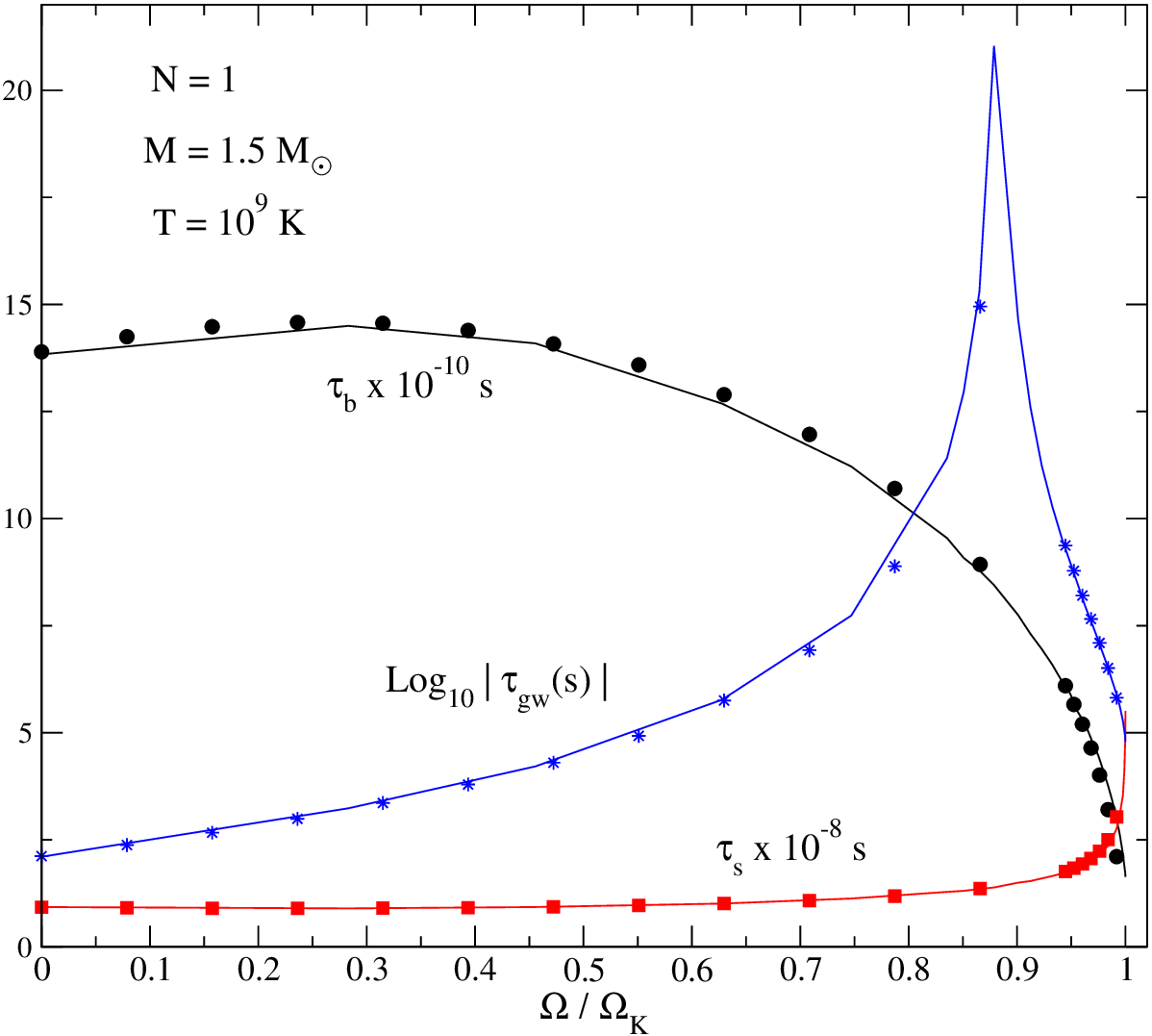}
\caption{ The left panel shows  the $l=m=4$ $f$-mode frequency (in the rotating frame) for 
our polytropic rotating models. The horizontal axis 
shows the angular velocity in the units of the  maximum rotation rate $\Omega_{\rm K}$.  
The right panel compares our results with previous work in literature, it displays 
the viscous and gravitational radiation damping/growth times of the $l=m=4$ $f$-mode 
for a sequence of polytropic rotating stars with  $N=1$, $M=1.5 M_{\odot}$ and  $T=10^{9}$~K . The 
vertical axis shows the various damping times in seconds (see legend). Our results (solid lines) 
and the data determined by~\citet{1991ApJ...373..213I,1995ApJ...438..265L}  (symbols) are in good agreement, especially 
in the CFS instability region of the $f$-mode. 
\label{fig2}}
\end{center}
\end{figure*}

\subsection{Boundary Conditions} \label{sec:BC}

The time evolution of non-axisymmetric oscillations requires the prescription of boundary conditions. 
We study  rotating stars with equatorial and rotation axis symmetry,  where the two-dimensional numerical 
grid extends over the region $0 \leq r/R(\theta) \leq 1 $ and $ 0\leq \theta \leq \pi / 2$. The quantity  $R(\theta)$ is the 
star's radius at a given latitude $\theta$. At each time step, we must then 
specify the variables at the surface, origin, rotational axis and equator.

At the origin~($r=0$) and the rotational axis~($\theta=0$), the perturbation variables  and equations must be regular.  For the 
$m \ge 2$ non-axisymmetric modes the regularity condition leads to the following equations:
\begin{equation}
\delta \rho =  \delta h = \delta \Phi  = 0,  \qquad \textrm{and} \quad  \delta v^i = 0 \, ,
\end{equation}
both at $r=0$ and $\theta = 0$.

At the equator~($\theta = \pi/2$), the reflection symmetry divides the
perturbations  into two sets with opposite
parity~\citep{2009MNRAS.394..730P}. For the $l=m=4$ $f$-mode the parity conditions read
\begin{align}
& \frac{ \partial \delta \rho }{\partial \theta}  = \frac{ \partial \delta h }{\partial \theta}  = \frac{ \partial \delta \Phi }{\partial \theta}  = 0 \,  ,  \\
& \frac{ \partial \delta v^{r}}{\partial \theta}  = \delta v^{\theta} = \frac{ \partial \delta v^{\phi}}{\partial \theta} = 0   \, .  
\end{align}

We impose at the surface, $r=R(\theta)$,
the vanishing of Lagrangian enthalpy perturbation,
\begin{equation}
\Delta h = \delta h + \xi^{i}  \, \nabla_{i} h = 0 \, , \label{eq:Scond}
\end{equation} 
which is the standard free surface condition used in stellar oscillations.
The vector field $\xi^{i}$ is the Lagrangian displacement, and the value of the
perturbed enthalpy $\delta h$ at the surface
is  determined at each time step from equation~(\ref{eq:Scond}).

\subsection{Numerical tests} \label{sec:Tests}

We test our numerical framework using the results 
of~\citet{1990ApJ...355..226I, 1991ApJ...373..213I} and~\cite{1995ApJ...438..265L}.  These papers study 
  the $f$-mode  instability for various polytropic Newtonian models,  and use 
 an eigenvalue method to extract the frequencies and eigenfunctions of the $f$-mode.  
 
 First of all, we determine the frequency of the onset of 
  the CFS instability for inviscid stars. This corresponds to models with vanishing  
  $f$-mode  frequency  in the inertial frame, i.e. $\omega - m \Omega = 0$. We find 
$\Omega = 0.5604 \Omega_0$ for $N=1$ and 
 $\Omega =0.5498 \Omega_0$ for $N=3/4$. 
 These values agree to better than 0.5\% with the results reported in~\cite{1990ApJ...355..226I}. 
 Note that $\Omega_0= \sqrt{\pi G\bar \rho_0}$ is the frequency unit used by~\cite{1990ApJ...355..226I}, where 
 $\bar \rho_0$ denotes the average mass density of the non-rotating model. 
 In figure~\ref{fig2} (left panel) we show the $l=m=4$ $f$-mode frequencies determined with our numerical code.

 During the tests we have noticed a slight difference between the maximum rotation rate 
 determined with our numerical code and that  reported in~\citet{1991ApJ...373..213I}.
 Although the difference is very small (less than 0.6\%)  it has a visible impact on the instability window. 
  For the $N=1$ and $N=3/4$ polytropes, we find $\Omega_{\rm K} = 0.6352 \Omega_0$
  and $\Omega_{\rm K} = 0.6454 \Omega_0$, respectively, 
 while~\citet{1991ApJ...373..213I} report $\Omega_{\rm K} = 0.639 \Omega_0$ and $\Omega_{\rm K} = 0.648 \Omega_0$. 
We have no reason to question the reliability of our code, given that it reproduces to high accuracy the axisymmetric background configurations 
of~\citet{1986ApJS...61..479H}. Hence, we think that this small discrepancy is due to the different numerical techniques used  
in~\citet{1991ApJ...373..213I}.

  As a final test, we compare the damping/growth times  of the gravitational radiation and the viscous  damping times  
  with the data extracted from~\citet{1991ApJ...373..213I} and~\citet{1995ApJ...438..265L}. The 
  results are shown in figure~\ref{fig2}  for the $N=1$ rotating models   with temperature $T=10^{9}$~K.
  We have rescaled the data reported in~\citet{1991ApJ...373..213I} with the $\Omega_{\rm K}$ of our equilibrium configuration.
  By using the standard  bulk viscosity given in equation~(\ref{eq:saw}), we find a very good 
  agreement with previous results.

\begin{figure*}
\begin{center}
\includegraphics[height=75mm]{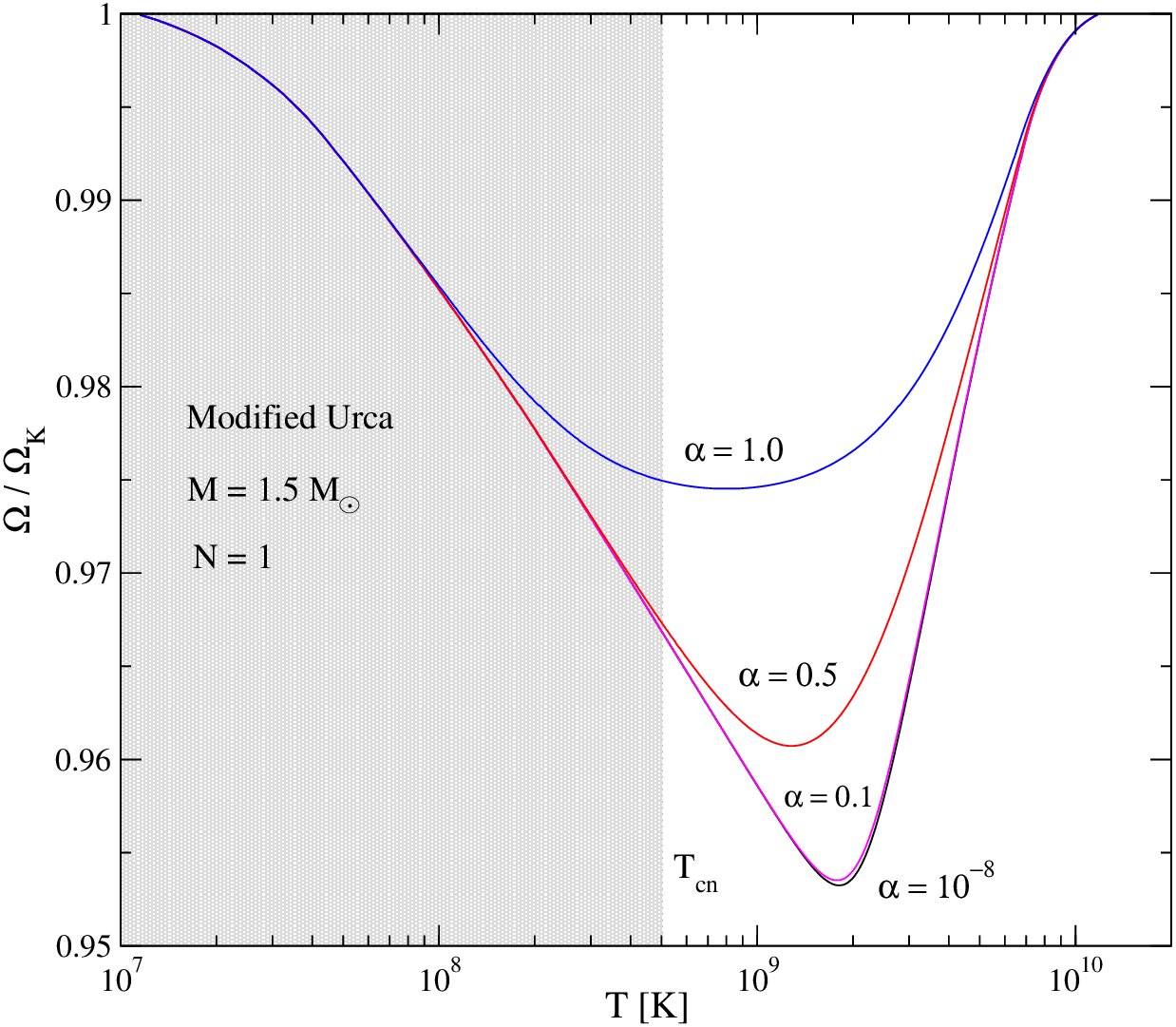}
\hspace{0.2cm}
\includegraphics[height=75mm]{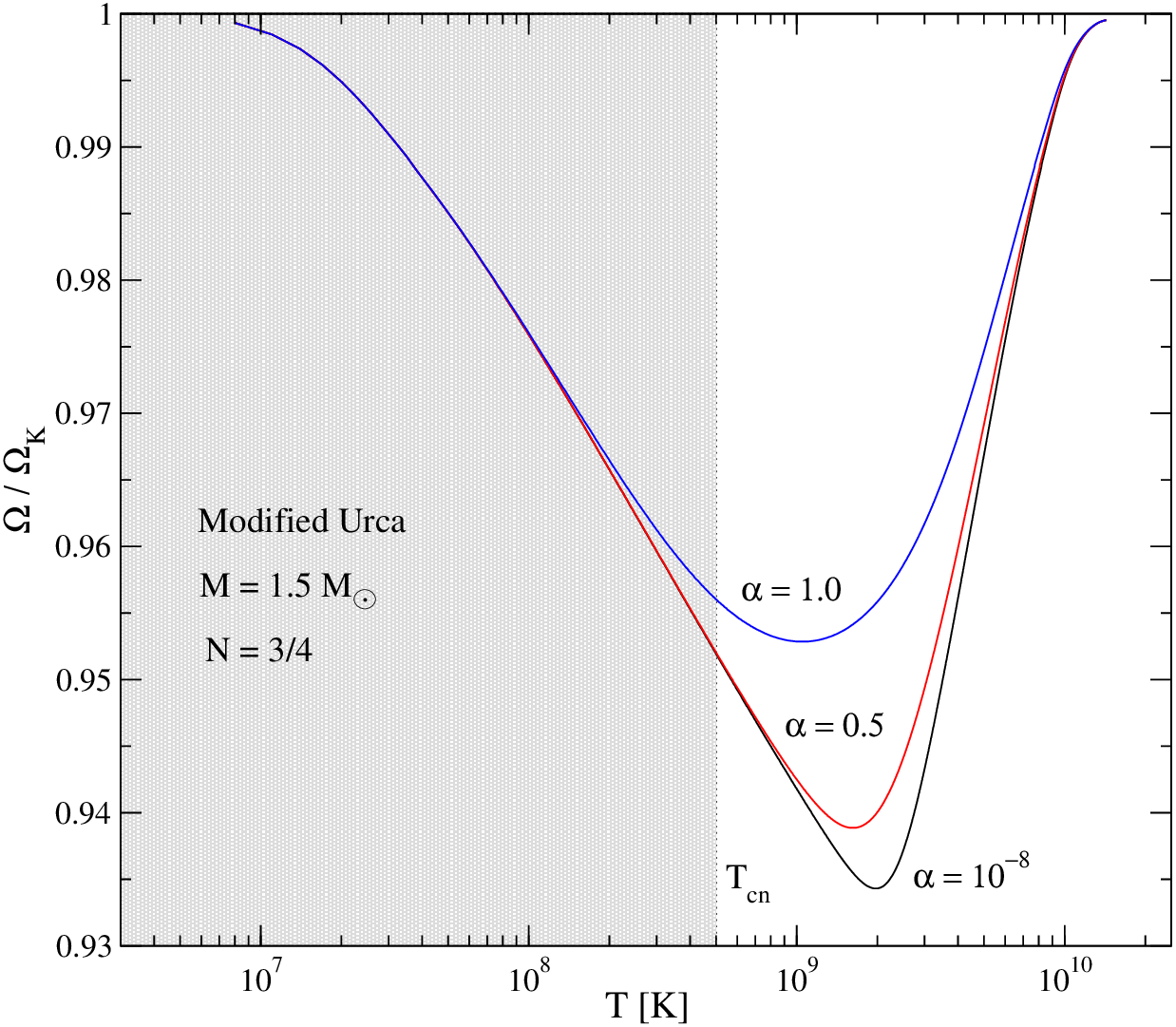}
\caption{
This figure displays the instability window of the $l=m=4$ $f$-mode 
for  the $N=1$ (left panel) and $N=3/4$ (right panel) polytropic models with mass $M=1.5~M_{\odot}$.  
The bulk viscosity is determined using the coefficient $\zeta$ (equation~(\ref{eq:zetaAl}))  
for the modified Urca process.   For various amplitudes  $\alpha$  we show 
the effects of the nonlinear bulk viscosity on the instability window.
The shaded regions represent the temperature range where neutron superfluidity (and proton superconductivity) in the core is 
expected to be present. The critical temperature $T_{\rm cn } \simeq 5\times 10^{8}$~K for the onset of neutron superfludity in the core represents 
a lower limit, as suggested by the recent models for the observed cooling of the Cassiopeia A neutron star 
~\citep{2011PhRvL.106h1101P,2011MNRAS.412L.108S}. Below that temperature superfluid mutual friction is expected to suppress the instability.
\label{fig3}}
\end{center}
\end{figure*}

\section{Results} \label{sec:results}

We are now ready to present our results on the impact of nonlinear bulk viscosity on the
gravitational wave-driven $f$-mode instability. We first discuss how the mode's instability window
is modified
as a result of viscous damping (section~\ref{sec:InsWin}). We then calculate the gravitational wave
strain associated with the $f$-mode and provide upper limits for its detectability  by present day and future
detectors (section~\ref{sec:GW}). In the final part of our calculation (section~\ref{sec:Rmode}) we make a brief digression and 
discuss the role of nonlinear bulk viscosity in the context of the $r$-mode instability.

\subsection{Instability window} \label{sec:InsWin}

The boundary of the $f$-mode instability window on the $\Omega-T$ plane is determined by
\begin{equation}
\frac{1}{\tau} = 0 \, . 
\label{eq:tauC}
\end{equation}
For a given $T$ this condition provides a critical rotational frequency  $\Omega_c$ above which
the mode is CFS-unstable. 

The previous studies of the $f$-mode instability included bulk viscosity in the sub-thermal regime, typically using
Sawyer's coefficient (\ref{eq:saw})~\citep{1989PhRvD..39.3804S}. In that case equation~(\ref{eq:tauC}), and consequently $\Omega_c$, 
are independent of the mode amplitude $\alpha$. However, once the full nonlinear bulk viscosity expression (\ref{eq:zetaAl}) is used, 
the instability window depends explicitly on $\alpha$. 

We first consider bulk viscosity due to the mUrca process. In figure~\ref{fig3} we show the $l=m=4$ 
$f$-mode instability window for our $N=1$ and $N=3/4$ polytropic models and for different mode amplitudes. The effect of
nonlinear bulk viscosity is clearly more pronounced in the $N=1$ model, and for amplitudes $\alpha > 0.1$. 
For an amplitude as large as $\alpha \sim 1$ the window area is decreased to a significant degree, resulting in a minimum $\Omega_c \approx 0.975 $.
The stiffer $N=3/4$ neutron star model is less sensitive to nonlinear viscous damping. This is not surprising given that the instability's 
growth time $\tau_{\rm gw}$ is shorter than the one in the $N=1$ polytrope. 
To  summarise, we can conclude that nonlinear bulk viscosity due to mUrca reactions has a {\em moderate} effect on the $f$-mode-instability
even for the largest mode amplitudes consistent  within our framework. 

\begin{figure*}
\begin{center}
\includegraphics[height=75mm]{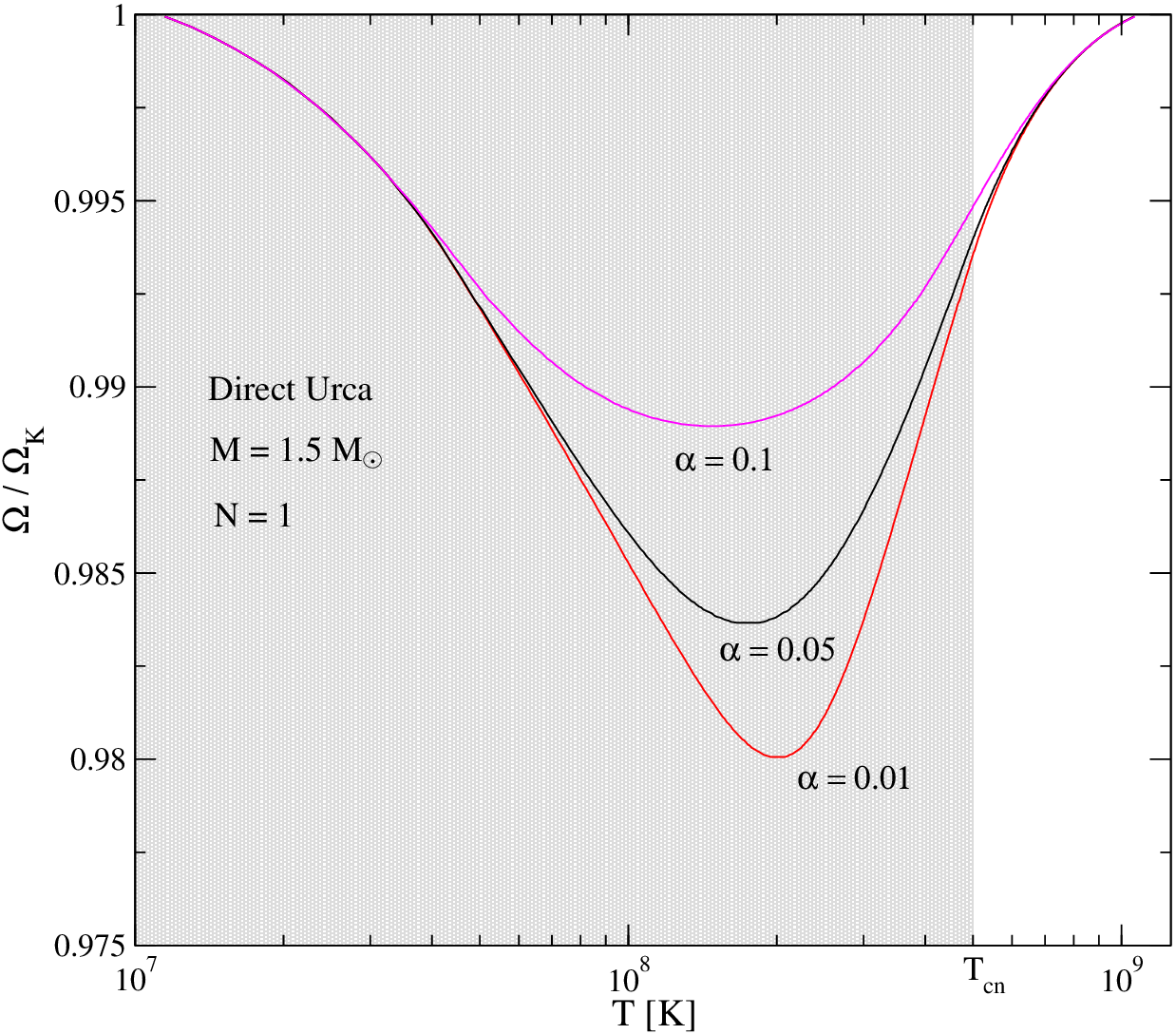}
\hspace{0.2cm}
\includegraphics[height=75mm]{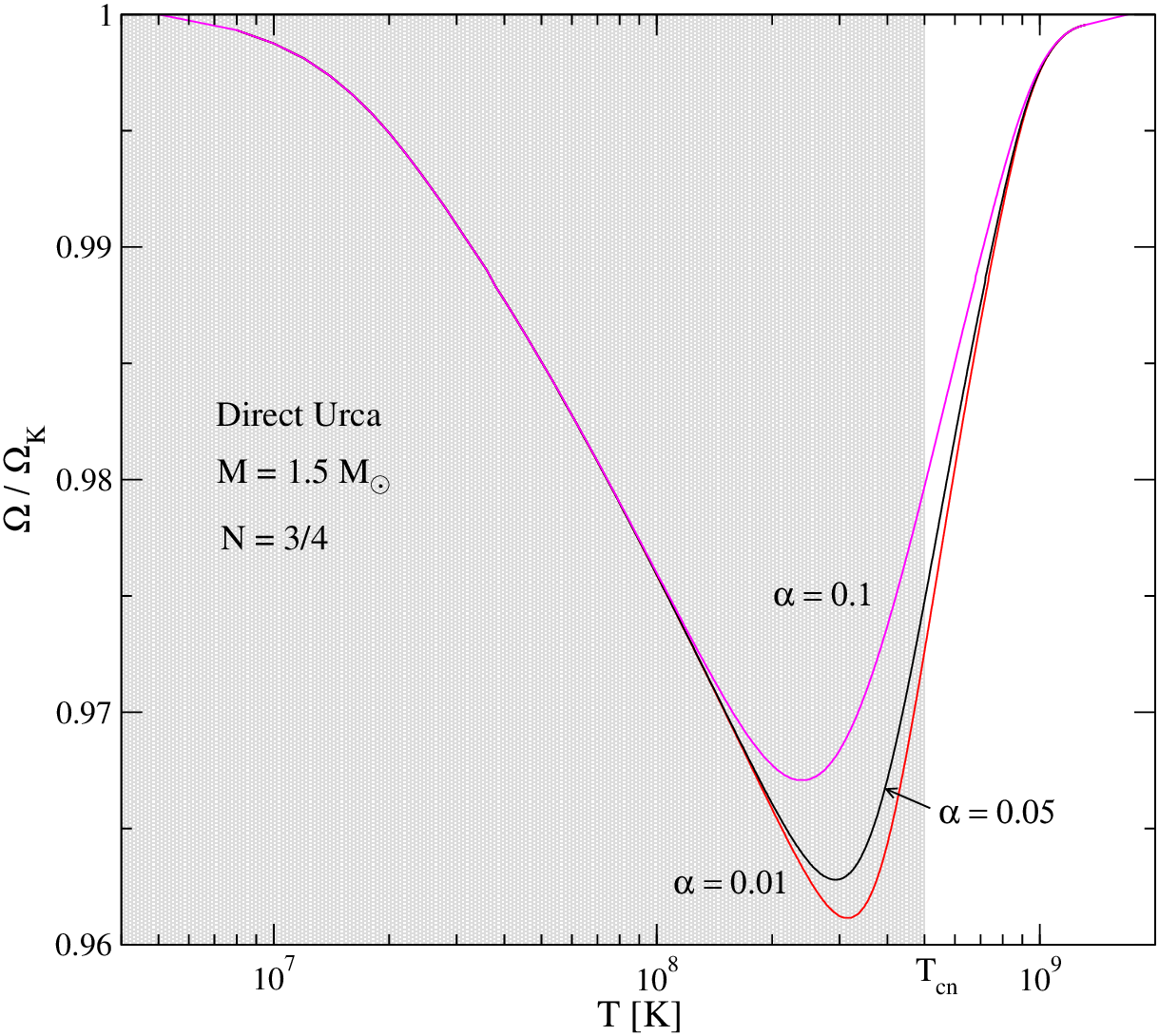}
\caption{This figure shows the instability window of the $l=m=4$ $f$-mode 
for  the $N=1$ (left panel) and $N=3/4$ (right panel) polytropic models with mass $M=1.5~M_{\odot}$.  
Differently from figure~\ref{fig3}, the bulk viscosity is here generated by the direct Urca process.
See figure~\ref{fig3} for more details about the notation.
\label{fig4}}
\end{center}
\end{figure*}

The instability window in models with dUrca reactions is shown in figure~\ref{fig4}. As expected, the higher efficiency of the dUrca reactions 
leads to a stronger bulk viscosity  which suppresses the instability above $T \approx 10^9\,\mbox{K}$. This temperature is very close to the expected 
onset of neutron superfluidity in the core, and once neutron superfluidity is present the instability is suppressed as a result 
of vortex mutual friction~\citep{1995ApJ...444..804L}. The resulting $f$-mode instability window is much smaller than the (already small) window 
of the mUrca models. Focusing on the damping of nonlinear bulk-viscosity we find that the situation is similar to the previous mUrca models. 
The critical frequency $\Omega_{c}$ is moderately increased, more for the $N=1$ polytrope and less for the $N=3/4$ one.

Our results suggest that bulk viscosity, despite its increased efficiency due to the nonlinear supra-thermal contribution, can only
cause a moderate degree of additional damping. How robust is this conclusion? Based on recent results on the $f$-mode
instability in relativistic stars~\citep{2011PhRvL.107j1102G} which generally predict a shorter growth timescale than the Newtonian
one ($\tau_{\rm gw}^{\rm GR} \sim 0.1 \tau_{\rm gw}^{\rm Newt}$), one would suspect that the role of nonlinear bulk viscosity is even less
important. At the same time, however, we also need to account for the fact that bulk viscosity itself becomes stronger, roughly by a factor
$\sim 10$, as we move to a more realistic model with interacting hadrons~\citep{2010JPhG...37l5202A, 2011arXiv1103.3521A}.
With these two effects combined and the ensuing balance between them (at least approximately) it is likely that our Newtonian analysis 
leads, after all, to robust results. Obviously, a definitive answer can only be reached by means of a more detailed calculation 
(e.g relativistic stellar models with a more realistic equation of state).  

%

As we have already pointed out, our analysis neglects the presence of neutron and proton superfluidity. 
At the same time, however, we know that superfluidity plays a key role by suppressing  the $f$-mode instability 
in sufficiently cold neutron stars through vortex mutual friction~\citep{1995ApJ...444..804L}. Bulk viscosity is also directly affected by superfluidity as a consequence 
of the reduced rates of the mUrca/dUrca reactions in superfluid neutron stars. Thus we have good reasons for revising 
the bulk viscosity damping in superfluid neutron stars. Within our framework we can do that by
introducing the appropriate superfluid correction factor in the viscosity coefficient $\zeta$ (see below). 
Other important aspects of superfluid hydrodynamics, like damping due to vortex mutual friction and other viscous degrees of 
freedom~\citep{2006CQGra..23.5505A}, 
can not be accessed without the full machinery of multi-fluid hydrodynamics~\citep*[e.g.][]{2009PhRvD..79j3009A}.

As a newborn neutron star cools down the transition to proton superconductivity is likely to occur first, with neutrons in the core 
becoming superfluid at a later stage. This order is the one suggested  by pairing calculations~\citep[e.g.][]{2002A&A...383.1076K}. 
The recent models for the observed thermal evolution of the Cassiopeia A neutron star provide constraints for the critical temperatures for the 
onset of proton and neutron superfluidity in the core, $T_{\rm cp} \approx (2-5) \times 10^9\,\mbox{K}$ and $T_{\rm cn} \approx (5-9) \times 10^8\,\mbox{K}$~\citep{2011PhRvL.106h1101P, 2011MNRAS.412L.108S}.
There is therefore a temperature range where the stellar core contains normal neutrons and superconducting protons. In this regime, where vortex
mutual friction cannot yet operate, it makes sense to study how bulk viscosity is modified by proton superconductivity. 

\begin{figure*}
\begin{center}
\includegraphics[height=75mm]{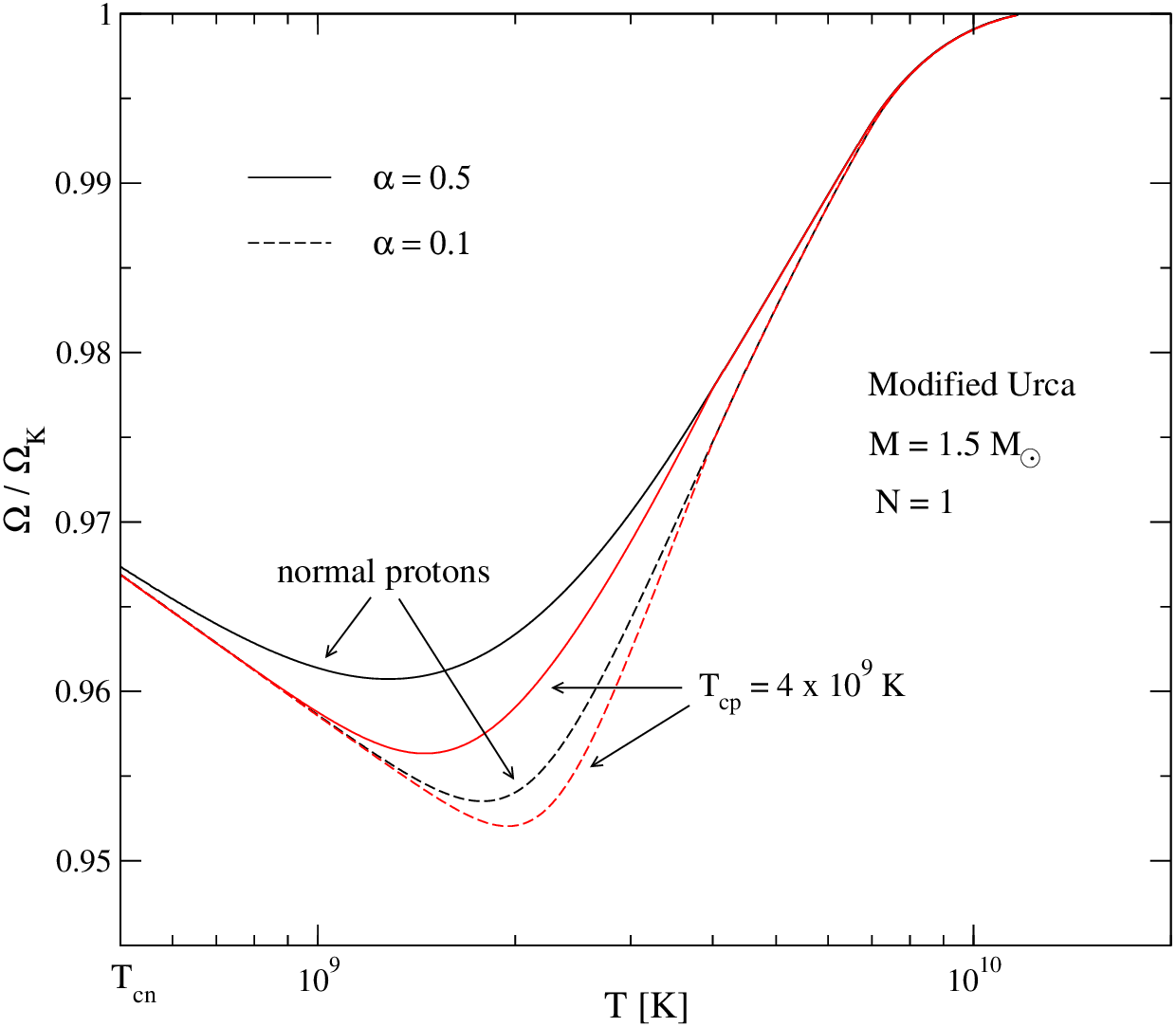}
\hspace{0.2cm}
\includegraphics[height=75mm]{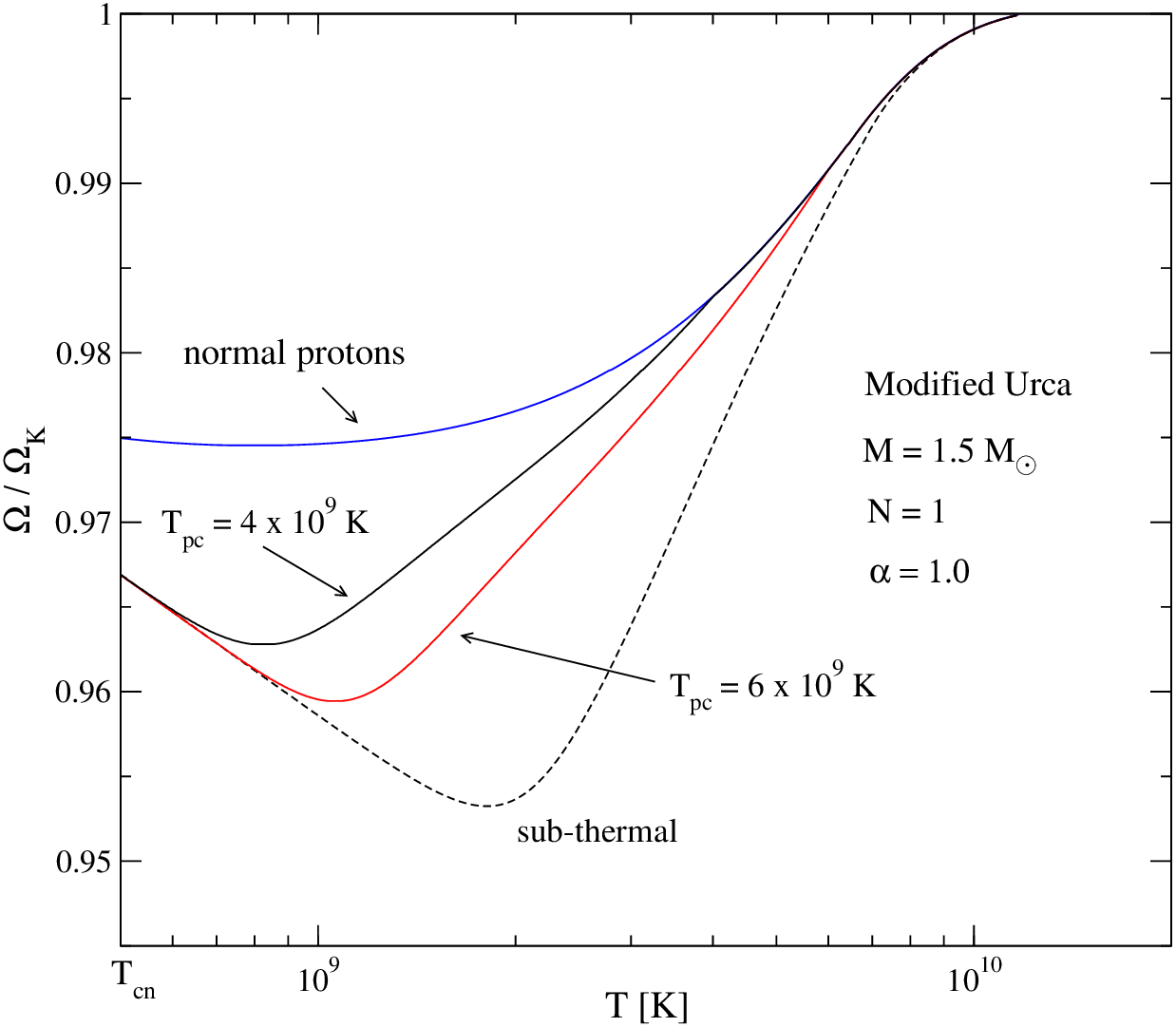}
\caption{
This figure shows the impact of proton superconductivity on the instability window of the $l=m=4$ $f$-mode 
for different choices of proton critical temperature $T_{\rm cp}$ (assumed to be uniform throughout the core).
The stellar mass is $M=1.5~M_{\odot}$ and the polytropic index $N=1$. The mode amplitude is $\alpha=0.1$ (dashed lines) and 
$\alpha=0.5$ (solid lines) in the left panel, and  $\alpha=1$ in the right panel. We show in the right panel also the critical curve for the 
sub-thermal bulk viscosity (dashed-line). 
\label{fig5}}
\end{center}
\end{figure*}

This can be achieved by using the modified viscosity coefficient 
\begin{equation}
\zeta_{\rm sup} = \zeta \, \mathcal{R} \, ,
\label{zeta_sf}
\end{equation}
where $\zeta$ is the bulk viscosity coefficient for normal matter  defined by equation~(\ref{eq:zetaAl}). 
The factor $\mathcal{R}$ represents the superfluidity-induced reaction rate suppression and is a function
of the critical temperature $T_{\rm cp}$ for the onset of proton superconductivity.
The explicit form of ${\cal R}$ is given by equation (34) in~\citet{2001A&A...372..130H}. It should be pointed out that
the modification (\ref{zeta_sf}) is the appropriate one for the bulk viscosity coefficient in the sub-thermal regime. 
As far as we know there is no similar result available for the supra-thermal $\zeta$, hence the superfluid modification (\ref{zeta_sf})
should be viewed as an approximation. 

The impact of proton superconductivity on the $f$-mode window is shown in figure~\ref{fig5}.  
We see that proton superconductivity  can  mitigate to a significant degree
the damping action of nonlinear bulk viscosity. This effect becomes more prominent with 
an increasing critical temperature $T_{\rm cp}$.
  
Nonlinear bulk viscosity, when efficient, is a mechanism for saturating a growing mode due to its intrinsic dependence
on the mode-amplitude. Of course this does not exclude the simultaneous presence of other competing (and perhaps more
efficient) saturation mechanisms. A strong candidate alternative saturation mechanism is nonlinear mode coupling.
For instance, mode couplings with other inertial modes is known to saturate the $r$-mode instability at small amplitudes, 
irrespectively of rotation/temperature~\citep{2003ApJ...591.1129A,2007PhRvD..76f4019B}. Thus it is also possible that
a similar effect could be at work in the case of the $f$-mode. Up to now there is limited work on the nonlinear
hydrodynamics of the $f$-mode~\citep{2004ApJ...617..490O, 2004PhRvD..70h4022S,  2010PhRvD..82j4036K}, and none for the $l=m=4$ multipole which is of interest to us. 

Still, we can obtain some hints from the available work on the quadrupole $f$-mode. The nonlinear numerical simulations 
of~\citet{2010PhRvD..82j4036K} suggest that the $f$-mode amplitude in relativistic polytropic stars is generically limited by wave breaking 
at the surface and mode couplings.  Considering a rotating $N=1$ polytropic model with axis ratio $R_p / R_{eq} = 0.7$, 
where $R_p$ is the polar radius,  
\citet{2010PhRvD..82j4036K} found that an $l=2$ $f$-mode with an initial energy 
$E \sim 2 \times 10^{-3} M_{\odot} c^2$ is efficiently damped by nonlinear effects on a timescale of about 10~ms. 
Assuming that  this result remains valid also for the  $m=4$ $f$-mode we can deduce (with the help of figure~\ref{fig1} and noting that
the above axis ratio corresponds to $\Omega \simeq 0.9 \Omega_{\rm K}$) that nonlinear mode couplings could saturate the mode amplitude at $\alpha < 1$.
This exercise suggests that nonlinear hydrodynamics may provide the dominant saturation mechanism for the $f$-mode instability. 
However, without a detailed analysis of the $l=m=4$ $f$-mode dynamics we cannot yet draw  any secure conclusion.


\subsection{Gravitational wave signal} \label{sec:GW}

The final part of our $f$-mode instability analysis is devoted to the  calculation of the gravitational wave strain 
 and its detectability. Unlike previous studies~\citep{1995ApJ...442..259L, 2004ApJ...617..490O, 2004PhRvD..70h4022S} 
we concentrate on the $l=m=4$ $f$-mode which is, as we have already pointed out, the most instability-prone 
multipole in both Newtonian and relativistic stars~\citep{2011PhRvL.107j1102G}. 

The gravitational wave strain can be computed by means of the standard multipole formula~\citep{1980RvMP...52..299T}
\begin{equation}
h_{ij}^{44} = \frac{G}{c^6} \frac{1} {r}  \frac{d^4
\mathcal{I}}{dt^4}^{44} \, T_{ij} ^{44} \, ,  \label{eq:hp22}
\end{equation}
where  $\mathcal{I}^{44}$ is proportional to the mass moment~(\ref{eq:Dlm}), 
\begin{equation}
\mathcal{I}^{44} = \frac{8 \pi }{945} \sqrt{5} \,  \delta D_{44}  \,  ,
\end{equation}
and $T_{ij}^{44}$ is the corresponding spin tensor harmonic angular function of
the ``electric-type''~\citep{1980RvMP...52..299T}. 

It turns out that the two independent wave polarizations can be expressed in terms of the 
spin-weighted spherical harmonic ${}_{-2}Y^{44}$ as
\begin{equation}
h_{\theta \theta}^{44} - i h_{\theta \phi}^{44} = h^{44} {}_{-2}Y^{44} \, , \label{eq:hdef}
\end{equation}
where  the wave amplitude $h^{44}$ is defined as  
\begin{equation}
h^{44} \equiv \frac{\sqrt{10}}{945}  \frac{4 \pi }{r}  \frac{G}{c^6}   \frac{d^4 }{dt^4}
\delta D_{44}   \,   .  \label{eq:h44b}
\end{equation}
For a monochromatic wave equation~(\ref{eq:h44b}) reduces to  
\begin{equation}
h^{44}  =  \frac{\sqrt{10}}{945}  \frac{4 \pi }{r}  \frac{G}{c^6}   \, \omega_I  ^4  \int d V\, \delta \rho( \mtb{r} )  \, r^4 Y_{44}^{\ast}    e^{i \omega_I t } 
= \hat h e^{i \omega_I t }  \,  ,  \label{eq:h44c}
\end{equation}
where $\omega_I = \omega -m\Omega$ is the mode frequency in the inertial frame. 

\begin{figure*}
\begin{center}
\includegraphics[height=75mm]{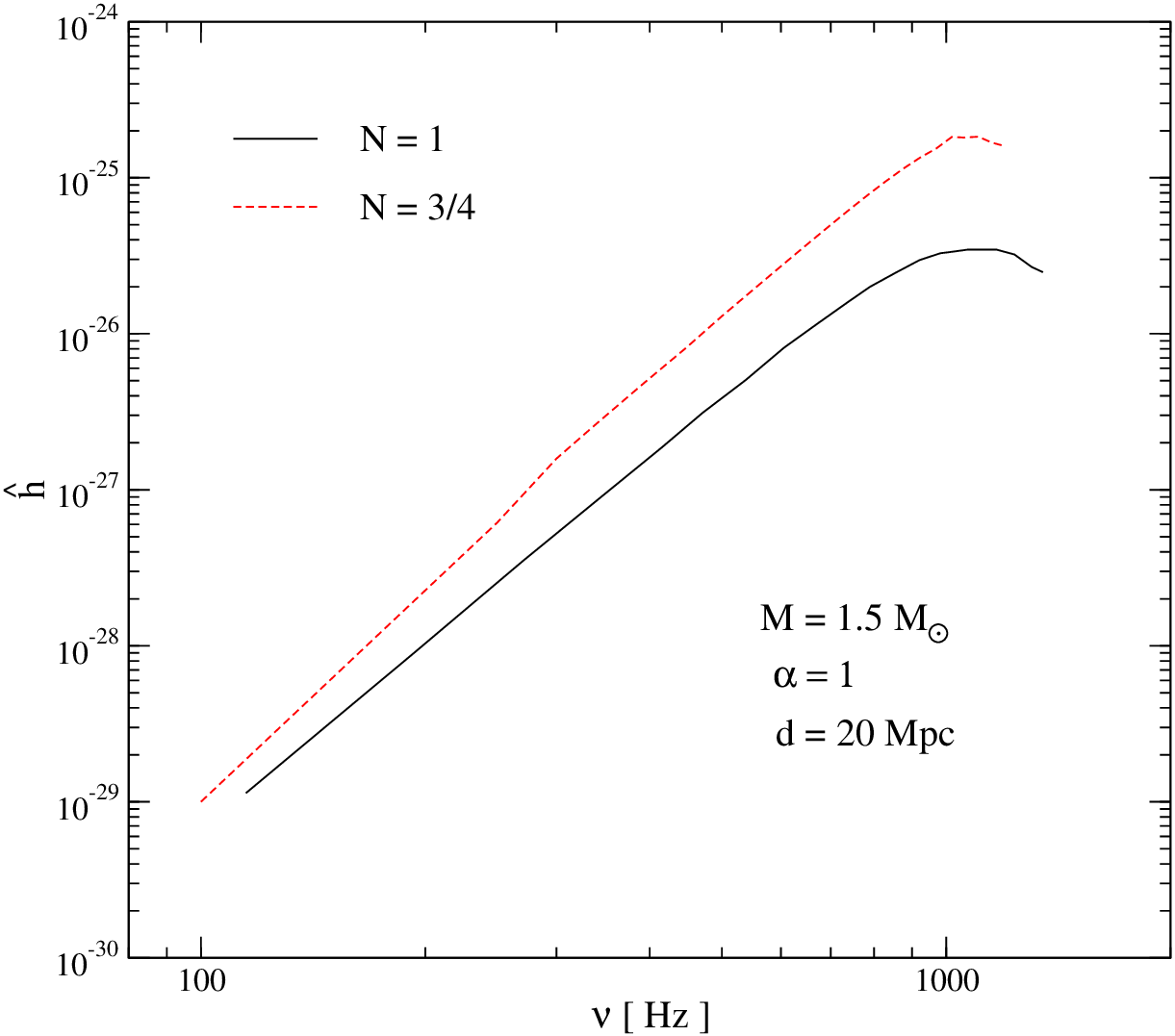}
\hspace{0.2cm}
\includegraphics[height=75mm]{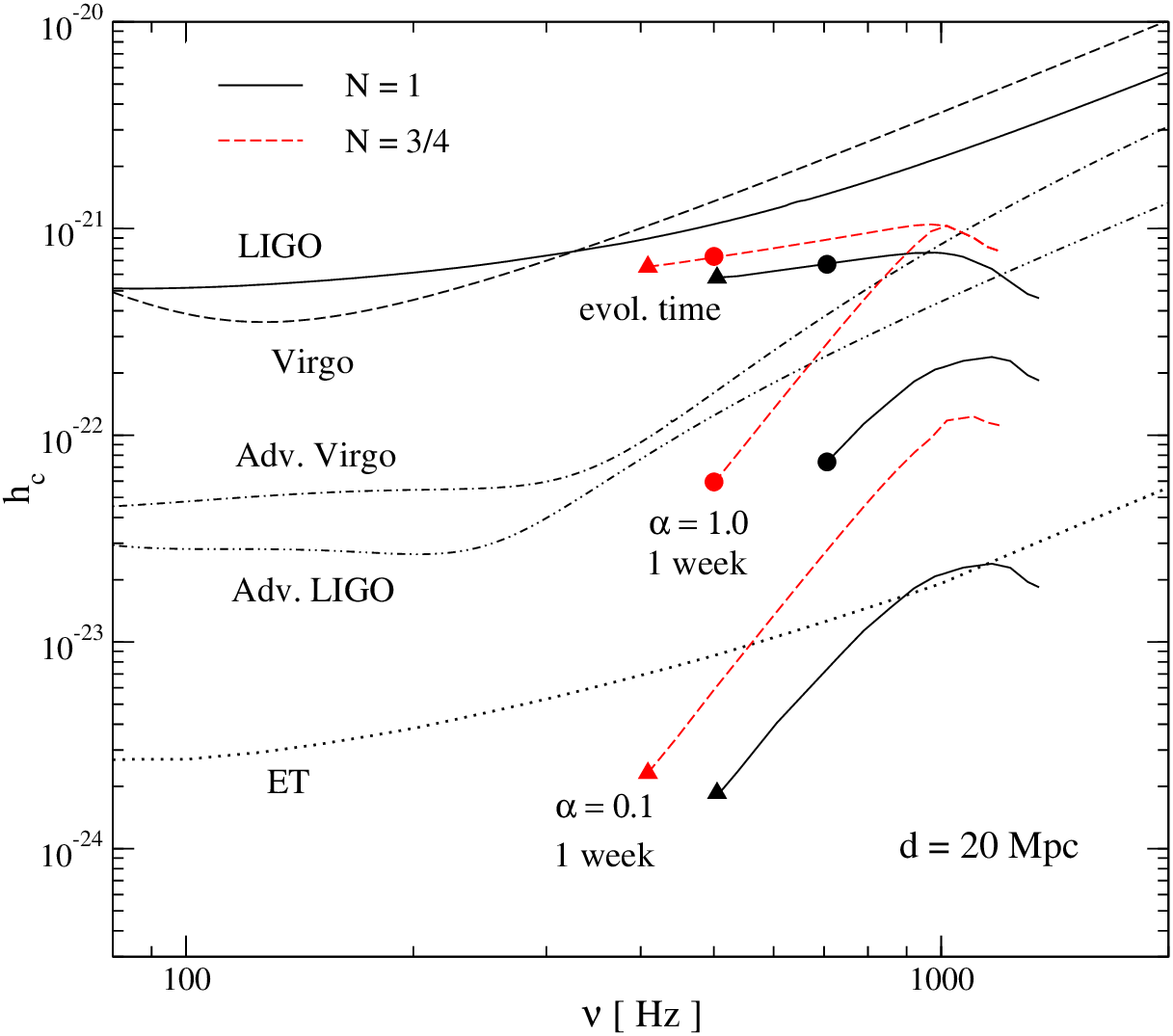}
\caption{Gravitational-wave strain generated by the unstable $l=m=4$ $f$-mode for rotating polytropic models with 
 indices $N=1$ and $N=3/4$. The stars have mass $M=1.5~M_{\odot}$ and are located at 20 Mpc. 
On the horizontal axis we show the inertial $f$-mode frequency, which  
can be directly related to the star's rotation rate (see the left panel of figure~\ref{fig2}). 
The left panel  displays the maximum strain $\hat h$ for an $f$-mode with fiducial amplitude $\alpha = 1$. 
In the right panel, we show the gravitational-wave signal emitted by the $f$-mode during the 
nonlinear saturation phase. 
We compare the characteristic strain $h_{c}$ for different mode amplitudes with the 
$h_{rms}$ of the gravitational-wave detectors (see legend). 
 The two curves with larger $h_c$ correspond to an  ideal case where the detectors can follow the entire evolution of the $f$-mode instability, and $h_{c}$ is then amplitude 
independent. The other curves represent a signal integrated for 1 week.  
Note that we have neglected the thermal evolution in our spin-down model, the curves in the right panel may therefore 
represent the gravitational radiation emitted by a star which slows down at constant temperature from the Kepler frequency to the 
minimum of the instability window. 
 For the mUrca process, we specify also the characteristic strain emitted by a model at the  minimum of the critical curve.  
 For an $f$-mode with  $\alpha =1$ ($\alpha=0.1$) this signal is denoted by filled circles (triangles).  
\label{fig6}}
\end{center}
\end{figure*}

As a representative gravitational wave amplitude we numerically calculate $h^{44}$ using the relevant $f$-mode
frequency and eigenfunctions. In figure~\ref{fig6} (left panel) we show our results for $\hat h$ assuming a source located at a distance
of 20~Mpc and a mode amplitude $\alpha=1$. As expected, the gravitational-wave signal is stronger for models with $\Omega \approx \Omega_{\rm K}$, 
where the star is  more unstable and  $\omega_{I}$ is higher.

For assessing the actual detectability of a neutron star undergoing oscillations due to an unstable $f$-mode we need
to take into account the cumulative effect of the observed number of oscillation cycles in a given frequency bandwidth.
This information is encoded in the characteristic wave strain
 \begin{equation} 
h_{c}  \equiv  \langle \left| h^{44} {}_{-2}Y^{44}  \right| \rangle \sqrt{  \nu^2 \left|  \frac{d t }{d \nu }  \right|  }  \label{eq:hc}
\end{equation}
where $\nu = \omega_I/2\pi$  and $\langle \dots \rangle$ 
denotes  the average over the angles $\left( \theta, \phi \right)$.

%

The time profile of the mode frequency can be determined following the two-stage evolution of the 
instability~\citep{1998PhRvD..58h4020O}. The onset of the instability leads to the first (linear) phase of exponential growth
where $\nu \approx {\rm constant}$. Eventually this growth arrives to a halt as the mode is saturated by nonlinear physics (for
instance, the nonlinear bulk viscosity studied in this paper). During this second (nonlinear) phase the mode amplitude
remains constant and the emitted gravitational radiation removes angular momentum from the star's bulk rotation causing its
spin-down. This evolution leads to a monotonically decreasing mode frequency $\nu(t)$. 
Here we  consider only the second nonlinear saturation phase and neglect the thermal evolution of the star and other mechanisms which can 
accelerate the spin-down. Our results provide therefore an approximate upper limit for the gravitational wave strain. 
A more detailed analysis of the detectability of unstable $f$-modes in neutron stars will be 
presented in future work.

In order to estimate the number of oscillations $N_{\rm cyc} = \nu^2 \left| d \nu / d t \right|^{-1} $
near the given mode frequency $\nu$ we need to consider the angular momentum balance for the system,
\begin{equation}
\frac{dJ}{dt} = - \left. \frac{dJ}{dt} \right|_{gw}  \, ,
\label{eq:Jbal}
\end{equation}
where $J=I\Omega $ is the stellar angular momentum \footnote{To be more precise we should also consider 
the variation of the mode's canonical angular momentum, but its contribution  is generally smaller 
than the stellar angular momentum~\citep{1998PhRvD..58h4020O} and hence can be neglected  for simplicity.}.
Writing $ dJ/dt = (dJ/d\Omega) (d\Omega/dt) $, $d\Omega/dt  = (d\Omega/d\nu) (d\nu/dt)$ and solving with respect to 
the derivative of the mode frequency, we obtain
\begin{equation}
\frac{d\nu}{dt} = - \frac{1}{2\pi} \frac{ d\omega}{d\Omega} \left. \frac{dJ}{dt} \right|_{gw} \left( \frac{dJ}{d\Omega} \right)^{-1} \,  .
\end{equation}
The radiated  angular momentum is given by~\citep{1980RvMP...52..299T}
\begin{equation}
\left. \frac{dJ_{i}}{dt} \right|_{gw} = 2 N_{4}   \, \omega_{I} ^{9} \left| \delta D_{44} \right| ^2 \delta_{i z}  \, .
\end{equation}
where $\delta_{i z}$ is the Kronecker delta. From this expression it is obvious the scaling $dJ/dt|_{gw} \sim \alpha^2$, which 
means that $N_{\rm cyc} \sim \alpha^{-2}$. Then from~(\ref{eq:hc}) we can deduce that the characteristic strain $h_c$ is 
independent of the mode amplitude $\alpha$.

The above calculation is idealised in the sense that it implicitly assumes a gravitational wave observation
lasting as long as the $f$-mode instability evolution timescale. This timescale is given by~\citep{1998PhRvD..58h4020O} 
\begin{equation}
\tau_{\rm{evol}} =  \left| \nu \frac{dt}{d \nu} \right| \, . 
\label{eq:tausd}
\end{equation}
For a fiducial mode amplitude $\alpha=1$ and a $N=1$ model,  we find that for $\Omega \approx \Omega_{\rm K}$ the evolution timescale is 
$\tau_{\rm{evol}} \sim 10$~weeks, and increases to $ \tau_{\rm{evol}}  \sim2$~yr for slower rotating models with $\Omega \simeq 0.98 \Omega_{\rm K}$. 
The evolution timescale becomes even longer for modes with $\alpha < 1$ given the scaling  $\tau_{\rm{evol}}\sim \alpha^{-2}$. 
Thus in the most relevant cases the mode evolution could be much longer than the typical observation time 
of LIGO and Virgo. What that means is that from the detectors' point of view the signal can be  considered 
as essentially monochromatic. Besides the temporal limits set by the detectors' capability, there might be additional ones from the physical system 
itself. For instance, the star may efficiently spin-down as a result of a strong magnetic field, thus quenching the $f$-mode
instability and the emission of gravitational radiation.

Therefore, in order to  provide a `realistic' estimate for the gravitational wave strain $h_c$ associated with the $f$-mode instability we
assume a monochromatic signal of duration $1\,\mbox{week}$, for a source at  a 20~Mpc distance. 
For the purpose of comparison we also consider
the ideal case of a signal lasting a time period $\tau_{\rm evol}$.  The results 
are shown in figure~\ref{fig6} together with the sensitivity curves of the current and future generation Earth-based gravitational 
wave detectors. The detectors' noise curves are computed in the standard way using $h_{\rm rms} \equiv  \sqrt{  \nu  S_{h} \left( \nu \right)  } $,
where $S_{h}\left( \nu \right) $ is the detector's power spectral density. 
 In our simplified spin-down model we have neglected the effects of the neutron star's cooling.  
The curves shown in figure~\ref{fig6} (right panel) may represent therefore the signal emitted by an unstable star 
which moves at constant temperature from the Kepler frequency down to 
the minimum of the instability window (see figure~\ref{fig3}).  
However the thermal evolution may significantly affect   
the trajectory  through the instability window and the star may therefore leave the instability region at larger rotation rates and thus  at  
higher  mode-frequencies than those indicated in figure~\ref{fig6}. 

Our results suggest that the $f$-mode instability may be 
detectable from the future ET for a mode amplitude $\alpha \ge 0.1$. Detection by Advanced LIGO/Virgo 
requires a larger amplitude $\alpha \sim 1$ and stars with a stiff equation of state (in the figure the $N=3/4$ polytrope is clearly
favoured over the $N=1$ one).

\subsection{Digression: the r-mode instability} 
\label{sec:Rmode}

So far we have discussed the role of (nonlinear) bulk viscosity in the damping of the $f$-mode instability. 
However, as mentioned in the introduction, the physics of nonlinear bulk viscosity has already been discussed
in the context of the $r$-mode instability~\citep{2010JPhG...37l5202A,2011AIPC.1343..580A,2011arXiv1103.3521A}.
The main conclusion of that recent work is that bulk viscosity operating in the
supra-thermal regime (see section~\ref{sec:bulk}) could suppress the $r$-mode instability (regardless of the stellar temperature), 
and prevent the growth of the mode amplitude above some maximum saturation level. 
In this section we revisit the topic of $r$-mode nonlinear viscous damping with the aim of providing a better understanding of the 
recent results. Why this is necessary will become obvious in the following discussion.  

To begin with, we introduce the `standard' dimensionless $r$-mode amplitude $\alpha$ defined as~\citep{1998PhRvD..58h4020O}
\begin{equation}
\delta v^j = \alpha \Omega R \left( \frac{r}{R} \right)^l Y^{Bj}_{lm} \, e^{i\omega t} \,  ,
\label{rmodedv}
\end{equation}
where $Y^{Bj}_{lm}$ is the magnetic vector harmonic. Obviously $\alpha$ must not be confused with the $f$-mode amplitude
defined in equation~(\ref{eq:norm}).
This expression describes the $r$-mode velocity at leading order with respect to the stellar spin 
in the familiar slow-rotation approximation~\cite[see][for further discussion]{2001IJMPD..10..381A}. 
In this approximation the star maintains its spherical shape and the mode's energy is dominated by the kinetic part, i.e. 
\begin{equation}
E = \frac{1}{2} \alpha^2  \Omega^2 R^{2-2m} \int_{0}^{R} \rho \,  r^{2m+2} dr \, , 
\end{equation}
where the $l=m$ $r$-mode is considered. 
The calculation of the bulk viscosity damping timescale requires the input of the density fluctuation associated with the perturbation 
(\ref{rmodedv}). 
After adding a missing factor in~\citet{1998PhRvD..58h4020O}~\cite[see also a comment in ][]{1999PhRvD..60f4006L}
\begin{align}
 \frac{\delta \rho }{\rho} & \approx \alpha  \frac{ \left( \Omega R\right)^2 }{c_s^2}  \frac{ 2 m +1}{m ( m + 1) \sqrt{ 2 m +3 } }   \nn \\ 
& \times \left[    \frac{2m}{2m+1} \sqrt{\frac{m}{m+1}} \left( \frac{r}{R} \right)^{m+1} + \delta \Phi_{0} \right]     Y_{m+1}^m \, e^{i\omega t}  \, , 
\label{eq:drh-rmode}
\end{align}
where $\delta \Phi_0 (r)$ is the radial eigenfunction of the perturbed gravitational potential. 
At this point it is important to make a clarification. In their study of the $r$-mode viscous 
damping~\citet{2011AIPC.1343..580A,2011arXiv1103.3521A}  have used a different 
definition for the $r$-mode amplitude. More specifically, Alford et al. use the amplitude $\tilde{\alpha}$ defined in equation (4.1) 
of~\citet{1999PhRvD..60f4006L}.\footnote{Shortly after the submission of this paper, Alford et al. 
have revised their r-mode calculation and used the more suitable amplitude defined by equation~(\ref{rmodedv}).}
The relation between the two different definitions of the $r$-mode 
amplitude is   \cite[see also the footnote in][]{1999PhRvD..60f4006L}
\begin{equation}
\alpha = \tilde \alpha  \sqrt{ \frac{\pi}{m}  \left( m + 1 \right)^3 \left(2 m +1 \right) !}    \, \, . \label{eq:aAl}
\end{equation}
For the most unstable $m=2$ $r$-mode we find $\alpha \simeq 71 \tilde \alpha$.

The fact that \citet{2011AIPC.1343..580A,2011arXiv1103.3521A}  use the notation $\alpha$ for the non-standard $r$-mode amplitude $\tilde{\alpha}$
may be the cause for some confusion. According to their results, e.g. fig. 2 in~\citet{2011AIPC.1343..580A}, 
bulk viscosity in the supra-thermal regime is capable of
suppressing the $r$-mode instability for a wide range of stellar temperatures provided $\tilde{\alpha} \gtrsim 0.05$. 
However, in terms of the standard $r$-mode amplitude (\ref{rmodedv}) this result corresponds to a saturation amplitude
$\alpha \gtrsim 3.5$. Obviously this is a rather large amplitude\footnote{This can be demonstrated if we take a $N=1$ 
polytropic star with mass $M=1.4M_{\odot}$ and radius $R=12.533$~km, rotating at the Kepler limit (the maximal rotation rate can 
be approximated with $\Omega_{\rm K} = \frac{2}{3} \sqrt{\pi G \bar \rho_0 }$, where $\bar \rho_0$ is the average density at $\Omega=0$). 
We then obtain $ E_{\rm kin} = 1.126 \, \alpha^2 \times 10^{51}\,\mbox{erg}$ and  $E_{\rm rot} = 1.8 \times 10^{52}\, \mbox{erg}$. 
Hence, the $r$-mode saturation amplitude obtained in~\citep{2011AIPC.1343..580A,2011arXiv1103.3521A} would imply a mode energy comparable to, 
or even larger, than the stellar rotational energy.}, much higher than the saturation amplitude  $\alpha \lesssim 10^{-3}$ due to 
the $r$-mode's nonlinear coupling with other inertial modes~\citep{2003ApJ...591.1129A,2007PhRvD..76f4019B}.  
Thus, we are led to conclude that for an amplitude $\alpha < 1$ the $r$-mode instability is affected very little by the nonlinear
contribution of bulk viscosity.

\section{Concluding remarks\label{conclusions}} \label{sec:concl}

We have studied the gravitational wave-driven instability of the $f$-mode in Newtonian polytropic neutron star models.  
The novelty of our work is the inclusion of bulk viscosity in the supra-thermal regime.
We have found that the ensuing nonlinear enhancement of bulk viscosity leads to a moderate change in the $f$-mode instability
window. For the particular case of the $l=m=4$ $f$-mode (the most unstable multipole) we have shown that nonlinear
bulk viscosity associated with the modified Urca process has a notable effect provided the mode amplitude, as defined in 
equation~(\ref{eq:norm}),  is  $\alpha \gtrsim 0.1$ (see figure~\ref{fig3}). This amplitude translates to a mode energy 
$E \gtrsim 10^{-5} E_{\rm rot}$ for $\Omega \approx \Omega_{\rm K}$, increasing to $E \gtrsim 10^{-3} E_{\rm rot}$ for $\Omega \approx \Omega_{c}$. 
 However, even for the extreme case of a mode amplitude as large as 
$\alpha \sim 1$ bulk viscosity cannot entirely suppress the instability.

If the more powerful direct Urca reactions can operate (which would be the case in the most massive neutron stars) then
the resulting $f$-mode instability window is significantly reduced already in the sub-thermal bulk viscosity regime.
The transition to the supra-thermal regime has again a moderate effect to the shape of the instability window. It turns out that 
in the temperature range where the change is important the instability is already likely to be suppressed by superfluid mutual
friction (see figure~\ref{fig4}).

The presence of superfluid components in neutron star matter could diminish the impact of bulk viscosity by suppressing
the rates of the chemical reactions involved. By considering a temperature regime where the protons are
superfluid and the neutrons normal and by treating in an approximate manner the superfluid suppression, we have shown 
that nonlinear bulk viscosity is significantly weakened  (see figure~\ref{fig5}).

Apart from calculating the bulk viscosity-modified $f$-mode instability window we have also provided upper limits for the
gravitational wave signal from a mode with $\alpha = 0.1-1$, i.e. the amplitude range where nonlinear bulk viscosity
can partially saturate the mode (figure~\ref{fig6}). Our results suggest that the $l=m=4$ 
$f$-mode instability (operating in a neutron star at the fiducial distance of 20~Mpc) could be detectable by a third generation 
instrument like the ET for an amplitude $\alpha \ge 0.1$ and an observation time of $\sim 1$ week. 
Detectability by Advanced LIGO/Virgo requires a rather high amplitude ($\alpha \sim 1$) and/or a longer observation time 
(e.g. the evolution timescale (\ref{eq:tausd})). Moreover, the $f$-mode  detectability is moderately sensitive to the 
stellar polytropic index, with lower indices (stiffer equation of state) being favoured. It should be pointed out that our 
work provides only `static' upper limits, in the sense that the combined rotational/temperature evolution of an $f$-mode unstable 
neutron star has not been properly incorporated. A more detailed analysis of this issue will the subject of future work. 

Our brief excursion to the study of the damping of the $r$-mode instability by nonlinear bulk viscosity has revealed that a 
large mode amplitude ($\alpha > 1$, see equation~(\ref{eq:aAl})) is required for the effect to be relevant. Thus, most likely, 
the saturation of the $r$-mode is controlled by nonlinear mode couplings rather than bulk viscosity. A similar statement could 
also be true for the $f$-mode but this remains to be (dis)proved, given the lack of any concrete results on the $f$-mode  
nonlinear couplings. Some light on this issue can be shed by a recent study of the nonlinear dynamics of the 
$l=2$ $f$-mode~\citep{2010PhRvD..82j4036K}. 
Extrapolation of  those results to the $l=4$ mode suggests 
that nonlinear hydrodynamics could outperform bulk viscosity in saturating a growing mode.

Besides the obvious need for a better understanding of the $f$-mode  nonlinear dynamics, several other improvements are
required in this research area. For instance, our own analysis can be clearly improved by moving from Newtonian to
General Relativistic gravity and from polytropic matter to more realistic equations of state. 
Indeed, as suggested by recent work on the relativistic $f$-mode, the instability's growth time may shrink as much as a factor $\sim 10$ 
compared to the Newtonian result. However, as we pointed out in section~\ref{sec:GW}, this effect is likely to be counter-balanced by a 
similar increase in the bulk viscosity strength as a result of using more realistic equations of state. The presence of a solid crust adds 
one more level of complexity to the problem. Despite its central role in limiting the $r$-mode 
instability~\citep{2001IJMPD..10..381A}, the dynamics of the crust has been totally ignored in the context of the $f$-mode instability.  
Another interesting possibility arises when exotic matter (for example, deconfined quarks) is present in the stellar interior. 
Previous work on the $r$-mode instability in neutron stars with exotic cores has unveiled several important differences with
respect to normal hadronic neutron stars~\citep[e.g.][]{1992PhRvD..46.3290M,2002MNRAS.337.1224A}. This is likely to be the case also for the $f$-mode instability.
Clearly, all the above issues are of great interest and we plan to address some of them in the near future.

\section*{Acknowledgements}
We thank L. Lindblom for providing us with the numerical data shown in figure~\ref{fig2}. We also would like to thank 
N. Andersson and K. Kokkotas for useful feedback on our paper, and W. Kastaun for fruitful discussions. 
AP acknowledges support from the German Science Foundation 
(DFG) via SFB/TR7. KG is supported by the Ram\'{o}n y Cajal Research Programme of the Spanish Ministerio
de Ciencia e Innovaci\'{o}n


\appendix

\nocite*


\label{lastpage}
\end{document}